\documentstyle[twoside,11pt,epsfig,color]{article}
\begin{document}

\title{Conformally Flat Collapsing Stars in f (R) gravity}

\author{Soumya Chakrabarti\footnote{email: soumya@cts.iitkgp.ernet.in}
\\
Centre for Theoretical Studies, \\Indian Institute of Technology, Kharagpur, \\West Bengal 721 302, India.
\\
Rituparno Goswami\footnote{Goswami@ukzn.ac.za}, Sunil Maharaj\footnote{Maharaj@ukzn.ac.za}
\\
Astrophysics and Cosmology Research Unit, \\School of Mathematics, Statistics and Computer Science, University of KwaZulu-Natal,\\ Private Bag X54001, Durban 4000, South Africa.
\\
Narayan Banerjee\footnote{email: narayan@iiserkol.ac.in}
\\
Department of Physical Sciences, \\Indian Institute of Science Education and Research, Kolkata, \\Mohanpur Campus, West Bengal 741246, India.
\date{}
}

\maketitle
\vspace{0.5cm}
{\em PACS Nos. 04.20.−q; 04.20.Jb; 04.50.Kd; 04.70.Bw
\par Keywords : gravitational collapse, exact solution, spherical symmetry, f(R) gravity}
\vspace{0.5cm}

\pagestyle{myheadings}
\newcommand{\be}{\begin{equation}}
\newcommand{\ee}{\end{equation}}
\newcommand{\bea}{\begin{eqnarray}}
\newcommand{\eea}{\end{eqnarray}}

\begin{abstract}
The present work includes an analytical investigation of a collapsing spherical star in $f(R)$ gravity. The interior of the collapsing star admits a conformal flatness. Information regarding the fate of the collapse is extracted from the matching conditions of the extrinsic curvature and the Ricci curvature scalar across the boundary hypersurface of the star. The radial distribution of the physical quantities such as density, anisotropic pressure and radial heat flux are studied. The inhomogeneity of the collapsing interior leads to a non-zero acceleration. The divergence of this acceleration and the loss of energy through a heat conduction forces the rate of the collapse to die down and the formation of a zero proper volume singularity is realized only asymptotically. 
\end{abstract}

\section{Introduction}
A modification of Einstein's gravity is one of the methods employed to explain the two phases of cosmic acceleration. Modification of the matter sector (for example, by including exotic scalar fields) is certainly the more popular approach to realize an effective negative pressure and therefore, the accelerated expansion of the universe \cite{inflation}. Modified gravity on the other hand modifies the action of General Theory of Relativity(GR) to write down an effective energy-momentum tensor with a geometrical origin.    \\

Perhaps the most straightforward modification of GR is understood by writing the Einstein-Hilbert lagrangian as an analytic function $f (R)$ of the Ricci scalar $R$ instead of $R$ itself. The so-called $f (R)$ theory of gravity can account for both the phases of cosmic acceleration, early in the history of the universe and the late-time accelerated expansion. A brief idea, from the motivations of studying the $f(R)$ modifications to the successes and shortcomings of the theory can be found in the reviews by Clifton, Ferreira, Padilla and Skordis \cite{cliff1}, Sotiriou and Faraoni \cite{soti1}, Nojiri and Odintsov \cite{nojiodirev1}, Nojiri, Odintsov and Oikinomou \cite{nojiodirev2}. Many specific choices of $f(R)$ have been put forward over the years, to address questions posed by increasingly advanced observational cosmology, or simply from a mathematical curiosity for a 'more general' theory of gravity. An $f(R)$ model is expected to describe a correct cosmological dynamics, without compromising the stability of its own. It should trace a precise newtonian and post-newtonian limit and cosmological perturbations well suited with the cosmic microwave background and large-scale structure. Although these restrictions limit the freedom of choice of the functional form of $f(R)$, there are viable models present in literature which attempt to describe both the phases of cosmic acceleration. The specific case of $f (R) = R + \alpha R^{2} (\alpha > 0)$ carries additional importance since this was the very first model of inflation proposed \cite{starob}. \\

The present work targets a different problem of gravitational physics, that of Gravitational Collapse where these theories do pose substantial difficulties. Gravitational collapse is defined as the implosion of a massive stellar body which can not produce sufficient internal pressure to balance out the inward pull due to gravity. Unless forced to stabilize in some kind of dynamical equilibrium phase, the collapse can continue until a zero proper volume is reached. At a zero proper volume the curvature invariants, physical quantities like density and pressure are expected to diverge, producing a space-time singularity. The predictability of the collapse reaching a singularity at a finite future lies within the highly non-linear field equations. Specific case studies of gravitational collapse indicate that the ultimate singularity may or may not be hidden from a distant observer depending on initial collapsing profiles. One defines these two possible outcomes of a collapse as, a Black Hole where the singularity is hidden behind a horizon, or a Naked Singularity, where the singularity can communicate with an observer through exchange of photon. Collapse of a stellar object consisting of a fluid with no pressure was addressed for the very first time by Datt \cite{datt} and Oppenheimer and Snyder \cite{os}. Various aspects and implications of gravitational collapse was recently reviewed by Joshi \cite{joshi1, joshi2}. \\

However, the number of existing collapsing solutions in $f(R)$ gravity is rather limited. The most obvious reason for this is the fact that the field equations of $f(R)$ gravity are fourth order in metric components and an exact or approximate solution is not always guaranteed. The interest in $f(R)$ collapse models has picked up quite recently. Possibility of ending up in a finite time curvature singularity in a stellar collapse process was investigated by Bamba, Nojiri and Odintsov \cite{bamba} and Arbuzova and Dolgov \cite{arbuzho} for different $f(R)$ setup. A few recent investigations by Borisov, Jain and Zhang \cite{borisov} and Guo, Wang and Frolov \cite{guo} demonstrated the utility of numerical simulations in a study of $f(R)$ collapse. The latter used a double-null formalism in Einstein frame, reducing the nonlinear curvature contribution into a nonminimally coupled scalar field by a conformal transformation. Kausar and Noureen \cite{kausar} studied the effect of pressure anisotropy and dissipation on the collapse dynamics for a specific $f (R)$ model written as $f(R) \sim R + \alpha R^n$. A case of charged and expansion-free $f(R)$ collapse was studied by Sharif and Yousaf \cite{sharif}. Very recently, Cembranos, Cruz-Dombriz and Nunez \cite{cembra1} studied uniformly collapsing cloud of self-gravitating dust particles in $f (R)$ gravity for a general $f(R)$.     \\

For a complete and consistent analysis of gravitational collapse, the collapsing interior must be matched smoothly with a proper exterior spacetime. This is a sector where an $f(R)$ collapse model can pose additional difficulty. The fourth order field equations produce extra constraint equations which must be satisfied at the boundary hypersurface as well as the usual Israel-Darmois \cite{israel, darmois} conditions of General Relativity. For a general scalar tensor theory, the general junction conditions were treated by Barrabes and Bressange \cite{barrabes}. The additional conditions in $f(R)$ gravity come from the matching of Ricci scalar and it’s normal derivative across the boundary. The importance of these additional constraints have been investigated only recently, by Deruelle, Sasaki and Sendouda \cite{deru}, Senovilla \cite{seno}, Clifton \cite{cliff3}, Clifton, Dunsby, Goswami and Nzioki \cite{cliff4}, Ganguly, Gannouji, Goswami and Ray \cite{ganguly}. Very recently Goswami, Nzioki, Maharaj and Ghosh \cite{goswamicollapse}, studied the dynamics of collapsing spherical stars in $f(R)$ gravity, for a specific case of Starobinski model ($f (R) = R + \alpha R^2$). Considering existance of pressure anisotropy and heat flux, they investigated the boundary matching with a vacuum exterior solution and observed that the collapsing solution mimics the Lemaitre-Tolman-Bondi solution of GR \cite{lemaitre, tolman, bondi}. Their work put forward the impossibility of matching a homogeneous collapsing star in $f(R)$ gravity to a static vacuum exterior spacetime across a fixed boundary surface. This makes existence of an Oppenheimer-Snyder-Datt \cite{datt, os} analogue of dust collapse model, i.e., a collapsing homogeneous dust ball with a vacuum exterior, inadmissible in $f (R)$ theories. Simple examples of collapsing model in $f(R)$ gravity discussed by Chakrabarti and Banerjee \cite{scnb1, scnbeuro} confirmed this result given by Goswami et. al. \cite{goswamicollapse}. Therefore, it remains very important to study models of collapsing solution in $f(R)$ gravity since many existing models become inconsistent due to the additional restrictions over the Ricci Scalar. \\

The present work deals with the collapse of a spherical star in $f(R)$ gravity addressing the additional constraints on the Ricci scalar, without any a priori specific choice of $f(R)$ and a general energy momentum distribution. The solution helps one to write the expressions of density, anisotropic pressure and the radial heat flux of the collapsing fluid, explicitly. However, the non-linearity of the system is taken care of by assuming the spacetime metric to be conformally flat. A very rapid collapsing nature of the spherical star is observed, which slows down with time. For different choices of $f(R)$, energy conditions are studied as well. One salient feature is that in spite of this being a simple model, the allowed inhomogeneity in $g_{00}$ begets a nontrivial acceleration, which is not there in the standard metric ansatz, such as the Tolman-Bondi metric, which are normally taken up for the collapse models. Section $2$ and $3$ defines the action and the field equations of $f(R)$ gravity under conformal flatness. In section $4$, the interior spacetime is matched with a vacuum exterior solution and the collapsing behavior is discussed. Section $5$ treats the possibility of the formation of a horizon. In section $6$ and $7$, the time evolution and radial distribution of the physical quantities are presented, along with the time evolution of the energy conditions, for different choices of $f(R)$. We conclude the manuscript in section $8$.  

\section{$f(R)$ gravity in metric formalism}
The action of $f(R)$ gravity can be written as
\begin{equation}\label{action}
A=\int\Bigg(\frac{f(R)}{16\pi G}+L_{m}\Bigg)\sqrt{-g}~d^{4}x,
\end{equation}
where $L_{m}$ is the Lagrangian for the matter distribution. The variation of the action (\ref{action}) with respect to the metric tensor leads to the following fourth order partial differential equation,
\begin{equation}\label{fe1}
F(R)R_{\mu\nu}-\frac{1}{2}f(R)g_{\mu\nu}-\nabla_{\mu}\nabla_{\nu}F(R) + g_{\mu\nu}\Box{F(R)}=-8\pi G T^{m}_{\mu\nu},
\end{equation}
which defines the field equations in $f(R)$ gravity, where $F(R)=\frac{df}{dR}$.           \\

Writing this equation in the form of Einstein tensor, one obtains
\begin{equation}\label{fe2}
G_{\mu\nu}=\frac{\kappa}{F}(T^{m}_{\mu\nu}+T^{C}_{\mu\nu}),
\end{equation}
where
\begin{equation}\label{curvstresstensor}
T^{C}_{\mu\nu}=\frac{1}{\kappa}\Bigg(\frac{f(R)-RF(R)}{2}g_{\mu\nu}+\nabla_{\mu}\nabla_{\nu}F(R) 
-g_{\mu\nu}\Box{F(R)}\Bigg).
\end{equation}

$T^{C}_{\mu\nu}$ is technically the curvature contribution to the effective energy-momentum tensor. The effective energy-momentum distribution may or may not obey all the energy conditions like an ordinary fluid distribution.

\section{Conformal flatness and field equations}
We assume that the space-time metric has a vanishing Weyl tensor which indicates that it is conformally flat. Conformally flat radiating fluid spheres and shear-free radiating stars are very well-studied in literature \cite{conformal}. The metric can be written as

\begin{equation}
\label{metric}
ds^2={D(r,t)}^2\Bigg[dt^2-dr^2-r^2d\Omega^2\Bigg],
\end{equation}
where $D(r,t)$ is the conformal factor and governs the evolution of the sphere. The fluid inside the spherically symmetric body is assumed to be locally anisotropic along with the presence of heat flux. Thus the energy-momentum tensor is given by
\begin{equation}\label{EMT}
T_{\mu\nu}=(\rho+p_{t})u_{\mu}u_{\nu}-p_{t}g_{\mu\nu}+ (p_r-p_{t})\chi_{\mu}\chi_{\nu} + q_{\mu}u_{\nu} +q_{\nu}u_{\mu},
\end{equation}
where $q^{\mu}=(0,q,0,0)$ is the radially directed heat flux vector, $\rho$ is the energy density, $p_{t}$ the tangential pressure, $p_r$ the radial pressure, $u_{\mu}$ the four-velocity of the fluid and $\chi_{\mu}$ is the unit space-like four-vector along the radial direction. The vectors are normalized as
\begin{equation}
u^{\mu}u_{\mu}=1,\quad\chi^{\mu}\chi_{\mu}=-1,\quad\chi^{\mu}u_{\mu}=0.
\end{equation}

Inclusion of a non-zero pressure anisotropy may play a crucial role while dynamics of a massive collapsing star is considered. For discussions on these issues, we refer the reader to the work of Herrera and Santos \cite{herresan1}, Herrera et al. \cite{herresan2}. It was precisely proved by Herrera and Ponce de Leon that under the assumption of conformal symmetry, a smooth boundary matching with the exterior geometry is only possible if the fluid has a pressure anisotropy \cite{ponce}. A dissipative process in the form of heat conduction is there in the system as well. During a gravitational collapse, one expects the collapsing star to become increasingly compact over time. As the dimension of the component particles and the mean free path becomes more and more comparable, a dissipative process can help emit energy to settle the system down to an equilibrium, or slow down the collapse process such that the zero proper volume is reached only asymptotically. For more relevant details we refer to the work of Herrera et. al. \cite{herresan2}.

The four-acceleration is defined as $a_{\mu} = u_{\mu;\nu}u^{\nu}$ and the expansion scalar as $\Theta = u^{\beta}_{;\beta}$. For a comoving line element (\ref{metric}),
\begin{equation}\label{fourvel}
u^{\mu} = \frac{\delta^{\mu}_{0}}{D(r,t)}.
\end{equation} 

Using the equations (\ref{metric}) and (\ref{EMT}), the field equations (\ref{fe1}, \ref{fe2}) can be written for the present case as
\begin{equation}\label{ferho} 
3\frac{\dot{D}^2}{D^2}-2\frac{D''}{D}-4\frac{D'}{rD}+\frac{D'^2}{D^2} = D^2\Bigg[\frac{\rho}{F}+\frac{f}{2F}
-\frac{R}{2}\Bigg] +\Bigg[\frac{F''}{F}+\frac{\dot{D}\dot{F}}{DF}+\Bigg(\frac{2}{r}+\frac{D'}{D}\Bigg)\frac{F'}{F}\Bigg],
\end{equation}

\begin{equation}\label{fepr}
-2\frac{\ddot{D}}{D}+\frac{\dot{D}^2}{D^2}+4\frac{D'}{rD}+3\frac{D'^2}{D^2} = D^2\Bigg[\frac{p_{r}}{F} -\frac{f}{2F}+\frac{R}{2}\Bigg] + \Bigg[\frac{\ddot{F}}{F}-3\frac{\dot{D}\dot{F}}{DF}-3\frac{D'F'}{DF}-\frac{2F'}{rF}\Bigg],
\end{equation}

\begin{equation}\label{fept} 
-2\frac{\ddot{D}}{D}+\frac{\dot{D}^2}{D^2}+2\frac{D''}{D}+2\frac{D'}{rD}-\frac{D'^2}{D^2} = D^2\Bigg[\frac{p_{t}}{F} -\frac{f}{2F}+\frac{R}{2}\Bigg] +\Bigg[\frac{\ddot{F}}{F}-\frac{F''}{F}-\frac{\dot{D}\dot{F}}{DF}-\frac{D'F'}{DF}-\frac{F'}{rF}\Bigg],
\end{equation}

and

\begin{equation}\label{feq}
-2\frac{\dot{D}'}{D}+4\frac{\dot{D}D'}{D^2}= -\frac{qD^3}{F}+\frac{\dot{F}'}{F}-\frac{\dot{D}F'}{DF} -\frac{D'\dot{F}}{DF}.
\end{equation}

There are four independent equations but five unknowns, the components of the energy momentum tensor $\rho(r,t)$, $p_{r}(r,t)$, $p_{t}(r,t)$, $q(r,t)$ and the conformal factor $D(r,t)$. The straightforward way of studying solution of the field equations is to assume some form of restriction over the energy momentum tensor components, for example, pressure isotropy or vanishing pressure or a vanishing heat flux. Instead of specifying such an equation of state, we attempt to extract relevant information regarding the dynamics from the matching conditions of the collapsing sphere with a vacuum exterior described by the Schwarzschild solution.

\section{Matching with a vacuum exterior solution and an exact solution} 
If one assumes that the spherical collapsing star is surrounded by vacuum then the exterior spacetime is well described by a Schwarzschild geometry. Matching of the interior with a vacuum Schwarzschild exterior at a boundary hypersurface requires the continuity of both the metric and the extrinsic curvature on the boundary hypersurface, discussed extensively by Santos \cite{santosmatching} and Chan \cite{chanmatching}. Schwarzschild solution is given by

\begin{equation}\label{schwarz}
ds^{2} = \Big(1-\frac{2M}{R} \Big)dT^{2}- \Big(1-\frac{2M}{R} \Big)^{-1}dr^{2}-R^{2}(d{\theta}^{2}+\sin^{2}\theta d{\phi}^{2}),
\end{equation}
where $M$ is the total mass contained by the interior. For an interior metric given by 

\begin{equation}
\label{met-int}
ds^2=S^2(t,r)dt^{2}-N^2(t,r)dr^{2}-C^2(t,r)(d\theta^{2} +\sin^2\theta d\phi^{2}),
\end{equation}
the matching with the metric (\ref{schwarz}) yields (using the matching of the second fundamental form or the extrinsic curvature)
\begin{equation}\label{extrinscurv}
2\left(\frac{\dot{C'}}{C}-\frac{\dot{C}S'}{CS}-\frac{\dot{N}C'}{NC}\right)=^{\Sigma}
-\frac{N}{S}\left[\frac{2\ddot{C}}{C}-\left(\frac{2\dot{S}}{S} -\frac{\dot{C}}{C}\right) \frac{\dot{C}}{C}\right]+\frac{S}{N}\left[\left(\frac{2S'}{S}+\frac{C'}{C}\right)\frac{C'}{C}-\left(\frac{N}{C}\right)^2\right],
\end{equation}
where $\Sigma$ is the boundary. Also here we have used
\begin{equation}
R=^{\Sigma} C\;\;,\;\; Sdt =^{\Sigma} \sqrt{\Big(1-\frac{2M}{R} \Big)}dT 
\end{equation}

The Misner and Sharp mass function \cite{sharp}, defined as 
\begin{equation}
\label{MS-mf}
m(t,r)=\frac{C}{2}(1+g^{\mu\nu}C_{,\mu}C_{,\nu}),
\end{equation}
yields the mass contained by the surface defined by the radial coordinate $r$.    \\

For a conformally flat metric defined by (\ref{metric}), the condition (\ref{extrinscurv}) can be simplified into
\begin{equation}\label{ID}
2\frac{\ddot{D}}{D}+2\frac{\dot{D}'}{D}-\frac{\dot{D}^2}{D^2}-3\frac{D'^2}{D^2}-4\frac{\dot{D}D'}{D^2}-4\frac{D'}{Dr} {=}^{\Sigma} 0.
\end{equation}

Writing the equation (\ref{ID}) in terms of $A(r,t) = \frac{1}{D(r,t)}$, one gets
\begin{equation}\label{ID2}
2\frac{\ddot{A}}{A} + 2\frac{\dot{A}'}{A}-3\frac{\dot{A}^2}{A^2}-3\frac{A'}{Ar}+\frac{A'^2}{A^2} {=}^{\Sigma} 0.
\end{equation}

Let us assume that the boundary of the collapsing sphere is at a radial distance given by $r = r_{b}$. We also put in the ansatz that $A(r,t) = \eta(t) + \xi(r)$; $\eta(t)$ governs the time evolution of the collapsing sphere and $\xi(r)$ is a function of $r$ only, defined on the boundary hypersurface as
\begin{equation}
\xi(r_b) = \xi_{0},
\end{equation}

\begin{equation}
\xi'(r_b) = \xi_{1},
\end{equation}
and
\begin{equation}
\xi''(r_b) = \xi_{2}.
\end{equation}

Here a prime denotes a derivative with respect to $r$ and a dot represents a derivative with respect to $t$. With these assumptions, the term $\dot{A}' = 0$ and the equation (\ref{ID2}) simplifies considerably. At $r = r_{b}$, it can be written as
\begin{equation}\label{evol}
\frac{2\ddot{\eta}}{(\eta + \xi_0)} - \frac{3\dot{\eta}^2}{(\eta+\xi_0)^2} - \frac{3\xi_1}{r_b (\eta+\xi_0)} + \frac{\xi_{1}^2}{(\eta+\xi_0)^2} = 0.
\end{equation}
Equation (\ref{evol}) can be integrated to to yield a first integral as
\begin{equation}\label{evol2}
\dot{\eta}^2(\eta + \xi_0)^{-3} + \frac{3\xi_1}{2r_b}(\eta + \xi_0)^{-2} - \frac{\xi_{1}^2}{3}(\eta + \xi_0)^{-3} = \lambda.
\end{equation}
Here, $\lambda$ is a constant of integration. For all non-zero values of $\lambda$, it is difficult to integrate the equation (\ref{evol2})and write $\eta(t)$ in a closed form. We discuss the solution for the simplest case, i.e., for $\lambda = 0$. Since we have assumed $\eta(t)$ inversely proportional to the scale factor ($\dot{D} = -\frac{\dot{A}}{A^2}$), we note that for a collapsing solution, $\dot{\eta}$ must be positive all the time. Thus for $\dot{\eta} > 0$, equation (\ref{evol2}) is integrated to yield an exact solution written as
\begin{equation}\label{eta}
\eta(t) = \Big(\frac{2\xi_{1}r_{b}}{9} - \xi_{0}\Big) - \frac{3\xi_{1}}{8r_{b}}(t - t_{0})^2,
\end{equation}
where $t_{0}$ is a constant of integration.
Therefore the conformal factor $D(r,t)$ can be written as
\begin{equation}\label{exact}
D(r,t) = \frac{1}{\Big[\Big(\frac{2\xi_{1}r_{b}}{9} - \xi_{0}\Big) + \xi(r) - \frac{3\xi_{1}}{8r_{b}}(t - t_{0})^2\Big]}.
\end{equation}

From equation (\ref{exact}), it appears that the collapsing sphere, reaches zero proper volume only asymptotically in comoving time. However the proper time as calculated by $\tau = \int D(r,t)dt$ remains finite for any observer located at a given $r=r_0$. \\

Investigating the continuity of the Ricci scalar and its normal derivatives across the boundary hypersurface, we can comment on the nature of $\xi(r)$. The matching criterion of the Ricci Scalar was presented in details by Deruelle, Sasaki and Sendouda \cite{deru}. For more recent examples and discussions we refer to the works of Clifton, Dunsby, Goswami and Nzioki \cite{cliff4}, and Goswami, Nzioki, Maharaj and Ghosh \cite{goswamicollapse}.     \\

From the metric (\ref{metric}) and the exact solution (\ref{exact}), we write the Ricci scalar as
\begin{equation}\label{ricci}
R = \Bigg[\frac{9\xi_{1}\xi''}{8r_{b}}-\frac{81}{32}\frac{\xi_{1}^2}{r_{b}^2}+\frac{3\xi_{1}\xi'}{2rr_{b}}\Bigg] \Bigg[(t^{2} - 2tt_{0}) + \Lambda_{1}(r,\xi_{0},\xi_{1},\xi_{2}, t_{0}, r_{b})\Bigg].
\end{equation}

Here, the term $[\frac{9\xi_{1}\xi''}{8r_{b}}-\frac{81}{32}\frac{\xi_{1}^2}{r_{b}^2}+\frac{3\xi_{1}\xi'}{2rr_{b}}]$ governs the radial dependence of the scalar and time dependence is given by the term $[(t^{2} - 2tt_{0}) + \Lambda_{1}(r,\xi_{0},\xi_{1},\xi_{2}, t_{0}, r_{b})]$. The term $\Lambda_{1}(r,\xi_{0},\xi_{1},\xi_{2}, t_{0}, r_{b})$ works as a parameter consisting of $\xi$, $\xi'$, $\xi''$, $\xi_{0}$, $\xi_{1}$, $\xi_{2}$, $t_{0}$ and $r_{b}$ which we write in this form for the sake of brevity. For the continuity, the ricci scalar and it's spatial derivative must match at the boundary surface i.e.,
\begin{equation}
R \mid_{r_{b}} = R' \mid_{r_{b}} = 0.
\end{equation}
Then, the Ricci scalar can be written in a form like 
\begin{equation}\label{ricciform}
R = (r_{b}^2 - r^2)^2 g(t,r).
\end{equation}
Comparing equations (\ref{ricci}) and (\ref{ricciform}), we write
\begin{equation}
\frac{9\xi_{1}\xi''}{8r_{b}}-\frac{81}{32}\frac{\xi_{1}^2}{r_{b}^2}+\frac{3\xi_{1}\xi'}{rr_{b}} = (r_{b}^2 - r^2)^2.
\end{equation}
Writing this equation as a second order non-linear differential equation of $\xi(r)$, we get
\begin{equation}
\xi'' = \frac{8r_b}{9\xi_1}(r_{b}^2 - r^2)^2 - \frac{4\xi'}{3r} + \frac{9\xi_1}{4r_b}.
\end{equation}
The evolution of $\xi(r)$ can be written from this equation in an explicit form as
\begin{equation}\label{xi}
\xi(r) = \frac{4r_{b} r^6}{171\xi_1} - \frac{4r_{b}^3 r^4}{39\xi_1} + \frac{3}{14} r^{2} \Bigg(\frac{8r_{b}^5}{9\xi_1} + \frac{9\xi_1}{4r_{b}}\Bigg) - 3k_{1}r^{-\frac{1}{3}} +k_{2}.
\end{equation}
Here, $k_{1}$ and $k_{2}$ are arbitrary constants of integration. Differentiating equation (\ref{xi}) and putting $r = r_{b}$ gives a quadratic in $\xi_1$ as
\begin{equation}
\xi_{1}^2 - 28 k_{1} r_{b}^{-\frac{4}{3}} \xi_{1} - \frac{14}{5}r_{b}^{6} = 0,
\end{equation}
and therefore $\xi_{1}$ can be written as
\begin{equation}\label{quad}
\xi_{1} = 14 k_{1} r_{b}^{-\frac{4}{3}} \pm \sqrt{196 k_{1}^2 r_{b}^{-\frac{8}{3}} + \frac{14}{5} r_{b}^6}.
\end{equation}
To ensure that $\xi_{1}$ is real, one must ensure that 
\begin{equation}
k_{1}^2 + \frac{r_{b}^{\frac{26}{3}}}{70} > 0,
\end{equation}
which is valid for any real choice of $k_1$.  \\
From equation (\ref{exact}), it is straightforward to see that for a collapsing scenario ($\dot{D} < 0$), $\xi_{1}$ must be negative. From equation (\ref{quad}) it can be seen that this puts additional restrictions over $k_{1}$. We assume that $N_1$ and $N_2$ are the negative and the positive roots of the term under square root, respectively. Thereafter the necessary conditions that ensure a collapsing evolution are
\begin{equation}
k_{1} < -\frac{N_1}{14} r_{b}^{\frac{4}{3}},
\end{equation}
and
\begin{equation}
k_{1} > \frac{N_2}{14} r_{b}^{\frac{4}{3}}.
\end{equation}

\begin{figure}[h]
\begin{center}
\includegraphics[width=0.45\textwidth]{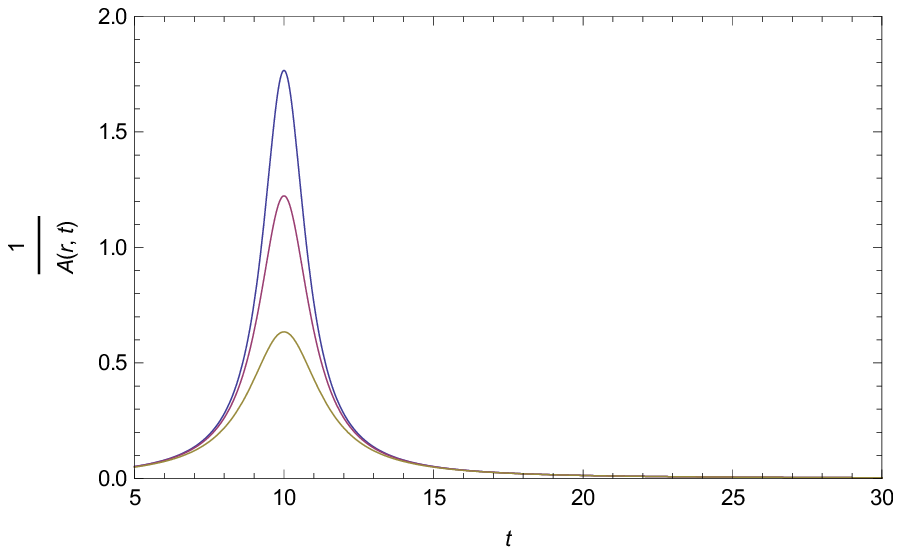}
\includegraphics[width=0.45\textwidth]{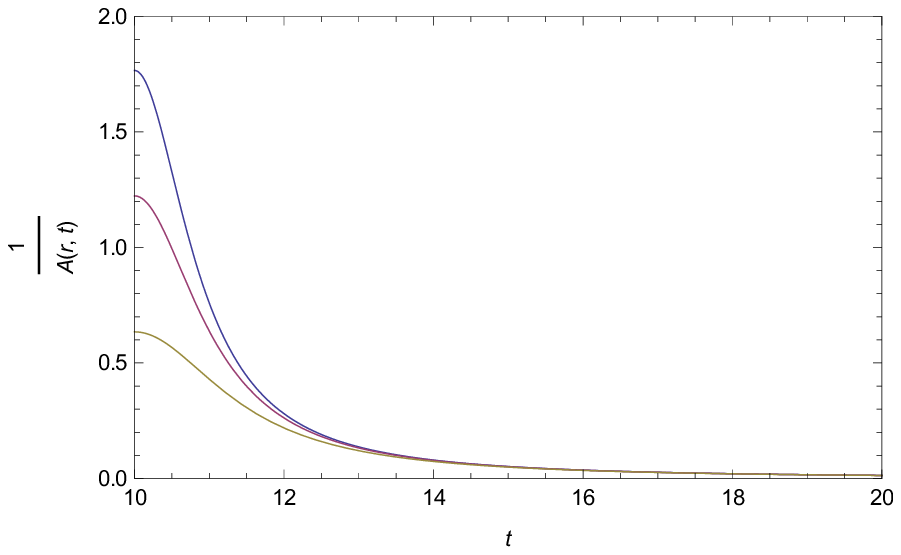}
\caption{\bf Time-evolution of the scale factor at different radial distance. The yellow graph shows the evolution at $r = r_{b}$. The red graph shows the evolution at $r = \frac{r_{b}}{2}$ and the blue graph $r = \frac{r_{b}}{10}$. The graph on the top indicates an initial brief period of expansion before the collapse begins.}
\end{center}
\label{fig:ltb}
\end{figure}

The time evolution of the spherical distribution is presented graphically in Figure $1$ for different values of $r$. It appears that, after an initial phase of expansion (depending on the constant of integration $t_{0}$, which can be put to $0$ without any loss of generlity), the sphere reaches a maximum accessible volume. Thereafter a rapid collapsing nature is prominent. Eventually the rapidity of the collapse diminishes with time and a zero proper volume is reached only asymptotically. Different curves show the nature of the collapse for different shells labelled by different values of the radial coordinate $r$. The shells start collapsing at a different rate, as in the innermost shell starts collapsing more rapidly, followed by subsequent shells of the fluid. However, with time the shells come closer and closer together and the curves denoting their collapse follow a more-or-less similar trajectory. The yellow graph (lowermost) shows the evolution at $r = r_{b}$. The purple graph (middle) shows the evolution at $r = \frac{r_{b}}{2}$ while the blue graph (top) is for $r = \frac{r_{b}}{10}$. \\

Now, using (\ref{fourvel}) and the metric (\ref{metric}), the four-acceleration and its' magnitude(scalar) can be written as
\begin{equation}\label{a1}
a_{1} = -\frac{D'(r,t)}{D(r,t)} = \frac{\Big(\frac{89}{42}r - \frac{8}{39}r^3 + \frac{4}{57}r^5\Big)}{\Big(\frac{5}{9}+\frac{89}{84}r^2-\frac{2}{39}r^4+\frac{2}{171}r^6+\frac{3}{4}(t-t_{0})^2\Big)},
\end{equation}
and
\begin{equation}\label{accelscalar}
a^{\mu}a_{\mu} = -\frac{{D'(r,t)}^2}{{D(r,t)}^4} = -\Big(\frac{89}{42}r - \frac{8}{39}r^3 + \frac{4}{57}r^5\Big)^2.
\end{equation}

We note that the magnitude of four-acceleration is a function of radial distance from the centre only. We plot $a^{\mu}a_{\mu}$ as a function of $r$ in Figure $2$.

\begin{figure}[h]
\begin{center}
\includegraphics[width=0.45\textwidth]{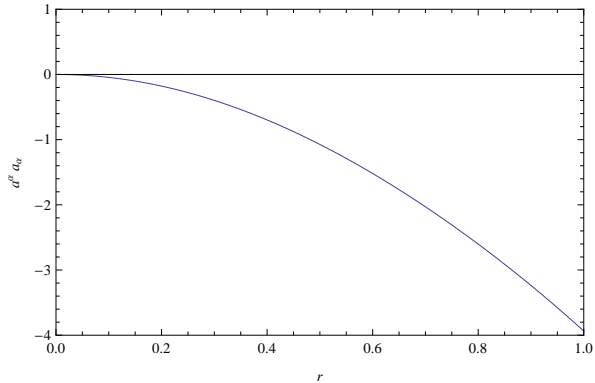}
\caption{\bf Evolution of the magnitude of four-acceleration as a function of r}
\end{center}
\label{fig:ltb}
\end{figure}

\section{Evolution of Density, Radial Pressure, Tangential Pressure and Heat Flux}
Since we have not assumed any specific equation of state, and the exact solution is obtained straightaway from a smooth boundary matching, the general expression of physical quantities like density, radial and tangential pressure, heat flux, can be written in terms of the metric components from the field equations (\ref{ferho}), (\ref{fepr}), (\ref{fept}) and (\ref{feq}). These features are all quite generic, as no specific functional form of $f(R)$ is assumed as yet.

\begin{figure}[h]
\begin{center}
\includegraphics[width=0.40\textwidth]{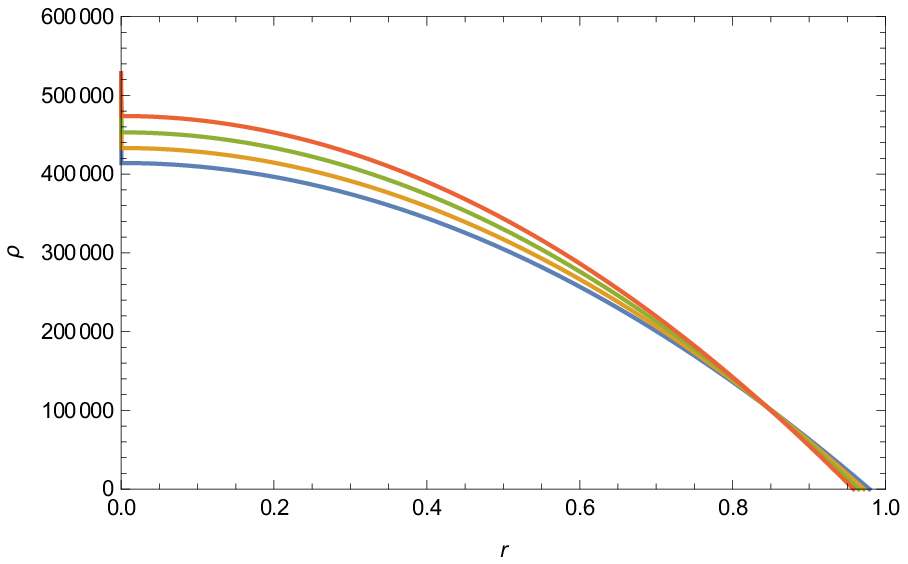}
\includegraphics[width=0.40\textwidth]{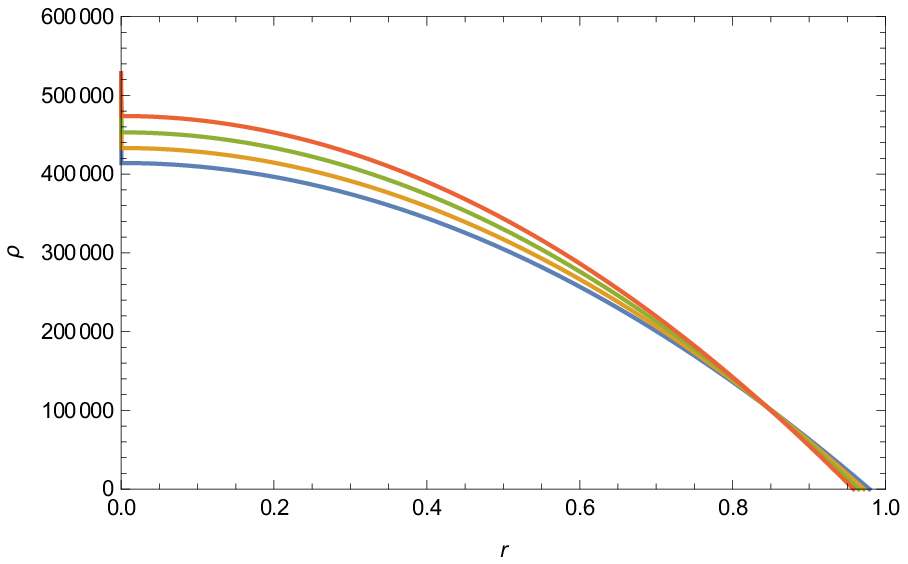}
\includegraphics[width=0.40\textwidth]{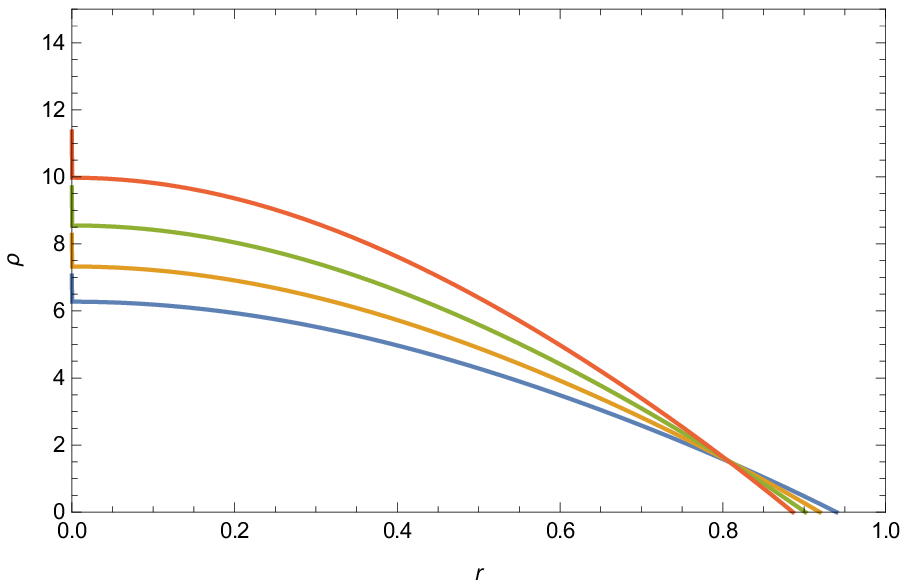}
\caption{\bf Evolution of the density with radial distance; $r=0$ to $r=1$ for different values of time for $f(R) = R + \alpha R^2$ (top left), $f(R) = R^2$ (top right) and for $f(R) = e^{\alpha R}$ (bottom). Different curves in a single graph denotes evolution for different values of time; the lowermost curve denotes evolution for $t = 10.0$, the curve second from the bottom denotes evolution for $t = 10.2$ and so on. The topmost curve denotes evolution for $t = 11.0$.}
\end{center}
\label{fig:ltb}
\end{figure}

In this section, we present a graphical analysis of physical quantities during the collapse for three popular and cosmologically important classes of $f(R)$ models; $f(R) = R^2$, $f(R) = e^{\alpha R}$ and $f(R) = R + \alpha R^2$, for different choices of the initial parameters. The value of the parameters governs nature of the evolution, however, the main qualitative nature remains similar. For $f(R) = R + \alpha R^2$, it is necessary to assume $\alpha$ to be small, otherwise the collapsing distribution doesn't match properly with a Scwarzschild solution at the boundary. We have worked with a fixed boundary hypersurface at $r_{b} = 1$ and $\alpha \sim 10^{-3}$ or lower, for $f(R) = R + \alpha R^2$, and with $\alpha \sim 0.16$ for $f(R) = e^{\alpha R}$. We keep other parameters fixed and plot the physical quantities as a function of $r$ for different values of time. Starting from $t= 10.0$, we plot upto $t=11.0$ at equal intervals of $0.2$ units of time.

We write the expression of density as
\begin{equation}
\rho = \Bigg[\Bigg(\frac{3\dot{D}^2}{D^2}-\frac{2D''}{D}-\frac{4D'}{rD}+\frac{D'^2}{D^2}-\frac{F''}{F}-\frac{\dot{D}\dot{F}}{DF}-\frac{2F'}{rF}-\frac{D'F'}{DF}\Bigg)\frac{1}{D^2}-\Bigg(\frac{f}{2F}-\frac{R}{2}\Bigg)\Bigg]F.
\end{equation}

\begin{figure}[h]
\begin{center}
\includegraphics[width=0.40\textwidth]{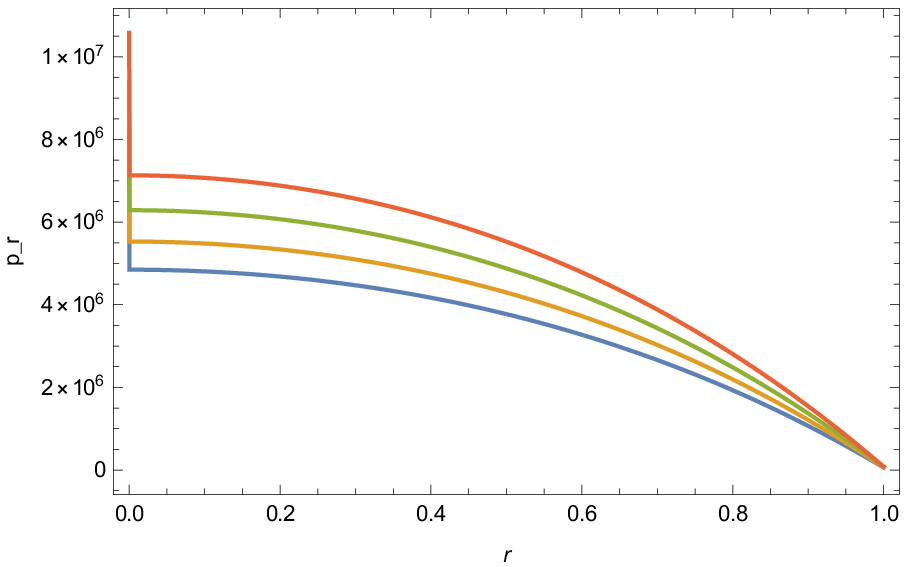}
\includegraphics[width=0.40\textwidth]{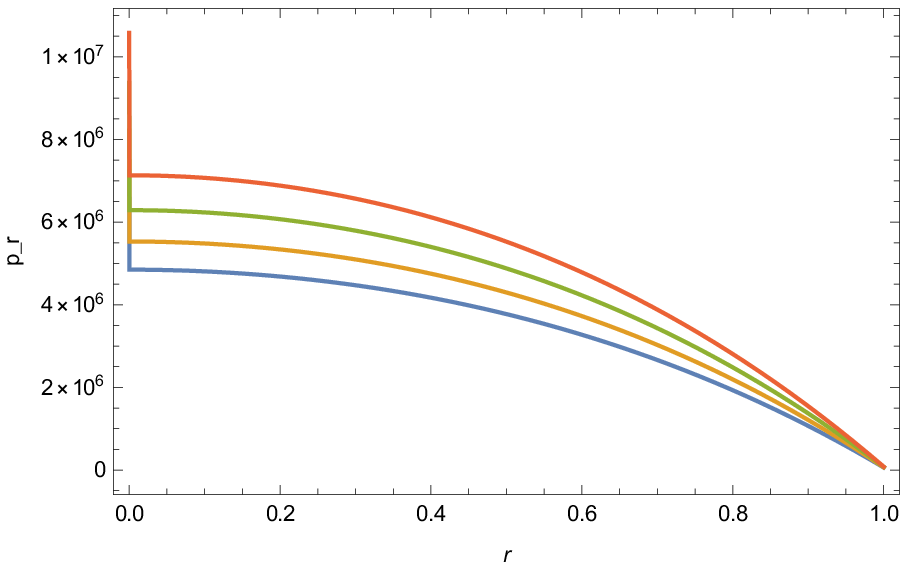}
\includegraphics[width=0.40\textwidth]{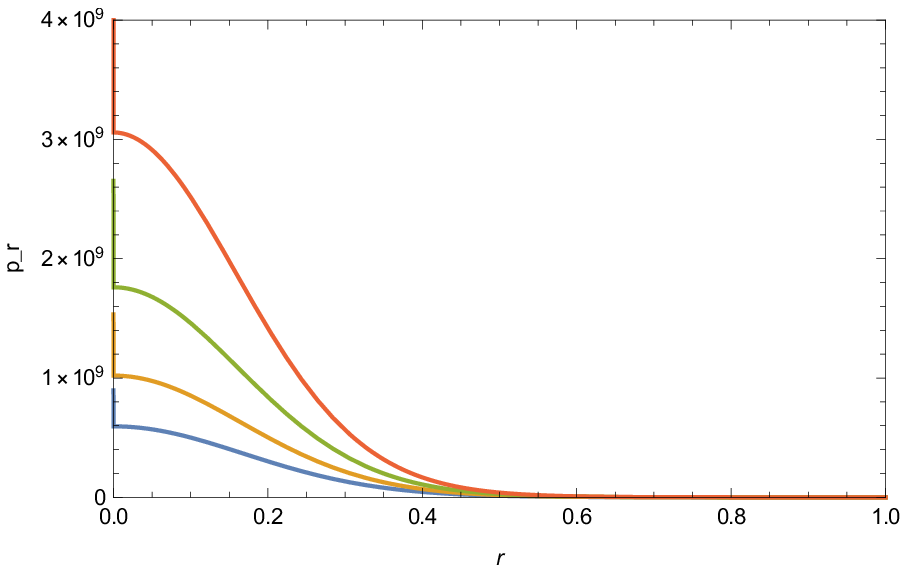}
\caption{\bf Evolution of radial pressure with radial distance; $r=0$ to $r=1$ for different values of time for $f(R) = R + \alpha R^2$ (top left), $f(R) = R^2$ (top right) and for $f(R) = e^{\alpha R}$ (bottom). Different curves in a single graph denotes evolution for different values of time; the lowermost curve denotes evolution for $t = 10.0$, the curve second from the bottom denotes evolution for $t = 10.2$ and so on. The topmost curve denotes evolution for $t = 11.0$.}
\end{center}
\end{figure}

In Figure $3$, we plot the distribution of density of the collapsing sphere as a function of the radial distance for different epoch. The lowermost curve shows the distribution when $t = 10.0$ and the uppermost one is for $t = 11.0$. Gradually, the density increases with radial distance decreasing, being the maximum at the centre ($r = 0$). Therefore, it is clear that the collapsing matter keeps on piling up at the centre and the core density increases with time. The density is strictly positive throughout the collapse and it becomes zero around the boundary hypersurface.

\begin{figure}[t]
\begin{center}
\includegraphics[width=0.40\textwidth]{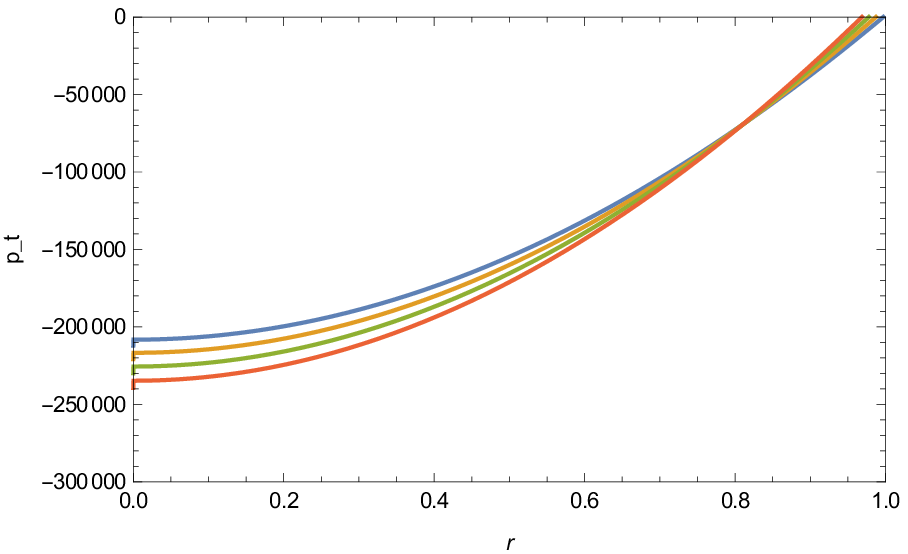}
\includegraphics[width=0.40\textwidth]{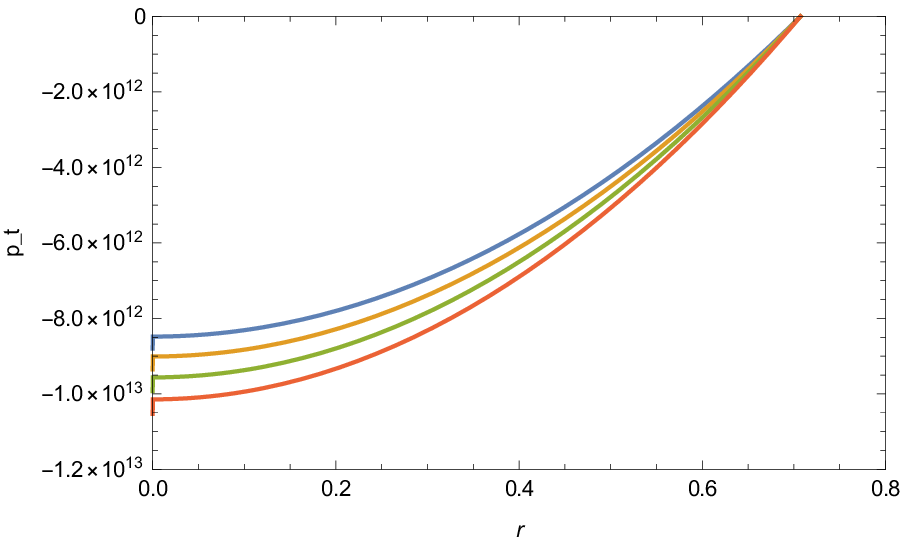}
\includegraphics[width=0.40\textwidth]{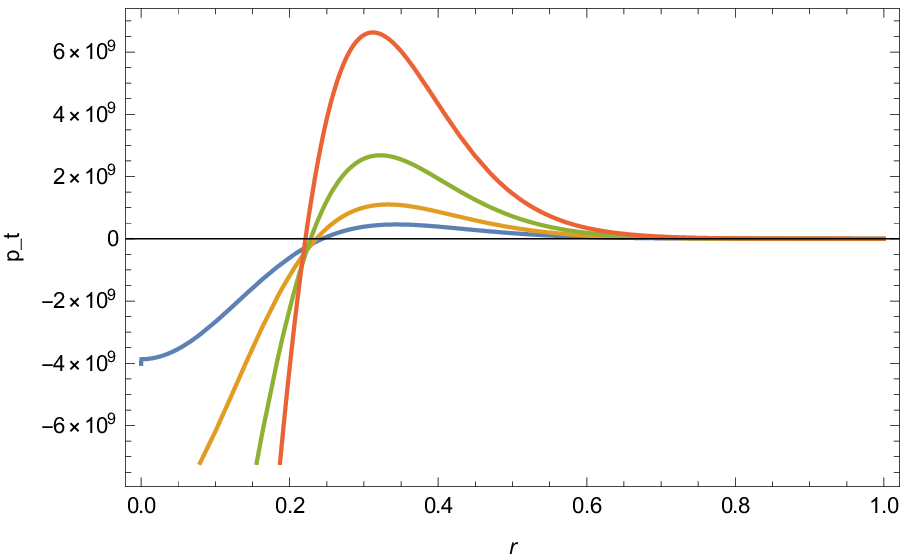}
\caption{\bf Evolution of tangential pressure with radial distance $r=0$ to $r=1$ for different values of time for $f(R) = R + \alpha R^2$ (top left), $f(R) = R^2$ (top right) and for $f(R) = e^{\alpha R}$ (bottom). Different curves in a single graph denotes evolution for different values of time; the uppermost curve denotes evolution for $t = 10.0$, the curve second from the top denotes evolution for $t = 10.2$ and so on. The lowermost curve denotes evolution for $t = 11.0$.}
\end{center}
\end{figure}

Radial pressure is governed by the equation 
\begin{equation}
-2\frac{\ddot{D}}{D}+\frac{\dot{D}^2}{D^2}+4\frac{D'}{rD}+3\frac{D'^2}{D^2} = D^2\Bigg[\frac{p_{r}}{F} -\frac{f}{2F}+\frac{R}{2}\Bigg]+\Bigg[\frac{\ddot{F}}{F}-3\frac{\dot{D}\dot{F}}{DF}-3\frac{D'F'}{DF}-\frac{2F'}{rF}\Bigg].
\end{equation}

We study the distribution of radial pressure of the collapsing sphere as a function of the radial distance in Figure $4$. The radial pressure is strictly positive throughout the collapse. Gradually, the density increases with radial distance decreasing, being the maximum at the centre ($r = 0$). It becomes zero at the boundary of the star and increases at the centre, i.e. $r = 0$ as time goes. Different curves in a single graph denotes evolution for different values of time $t$. The lowermost curve shows the distribution when $t = 10.0$ and the uppermost one is for $t = 11.0$. One important point to note here, at $r \sim 0$, the radial pressure attains a very high but a finite value, which is out of range ($\sim 10^{10}$ or higher) of the graphs in Figure $4$.  \\
The evolution of tangential pressure is governed by the equation 
\begin{equation}
-2\frac{\ddot{D}}{D}+\frac{\dot{D}^2}{D^2}+2\frac{D''}{D}+2\frac{D'}{rD}-\frac{D'^2}{D^2} = D^2\Bigg[\frac{p_{t}}{F} -\frac{f}{2F}+\frac{R}{2}\Bigg]+\Bigg[\frac{\ddot{F}}{F}-\frac{F''}{F}-\frac{\dot{D}\dot{F}}{DF}-\frac{D'F'}{DF} -\frac{F'}{rF}\Bigg],
\end{equation}

We plot the tangential pressure as a function of $r$ in Figure $5$. The tangential pressure is negative throughout the collapse for $R + \alpha R^2$ and $R^2$. Different curves in a single graph denotes evolution for different values of time $t$. The uppermost curve shows the distribution when $t = 10.0$ and the lowermost one is for $t = 11.0$. The tangential pressure becomes zero at the boundary of the star and has the maximum value as one approaches the centre, i.e. $r = 0$. For $e^{\alpha R}$, as the lower graph suggests, the evolution with radial distance is negative near the centre of the collapsing star and zero at the boundary. However, as one approaches the boundary hypersurface from the centre, the evolution becomes positive from a negative profile. In this case as well, around $r \sim 0$, the tangential pressure attains a very high but a finite value, out of range of the graphs, as is clear from the lowermost graph of in Figure $5$.
 
\begin{figure}[h]
\begin{center}
\includegraphics[width=0.40\textwidth]{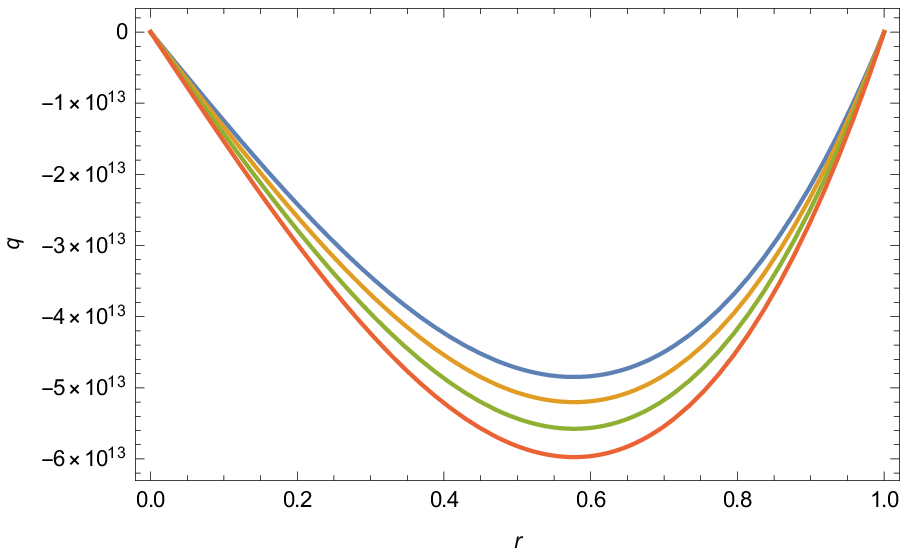}
\includegraphics[width=0.40\textwidth]{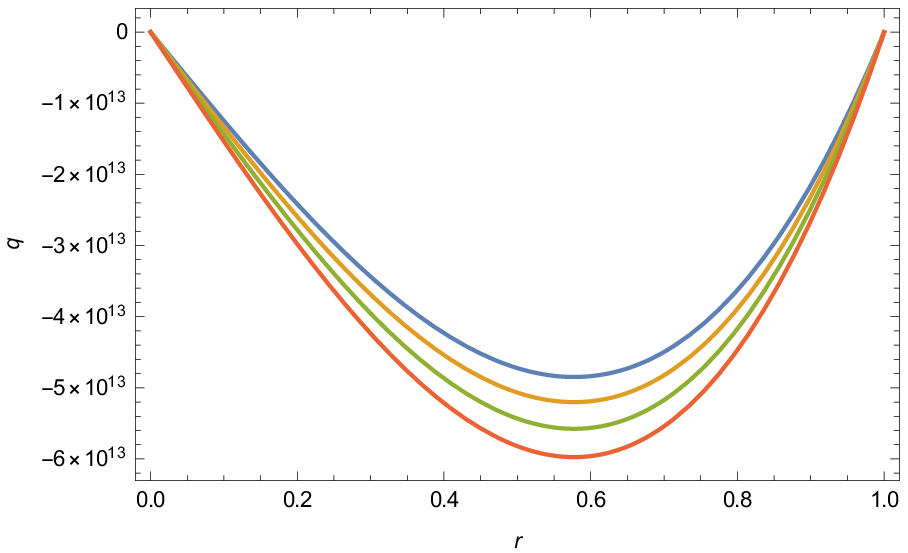}
\includegraphics[width=0.40\textwidth]{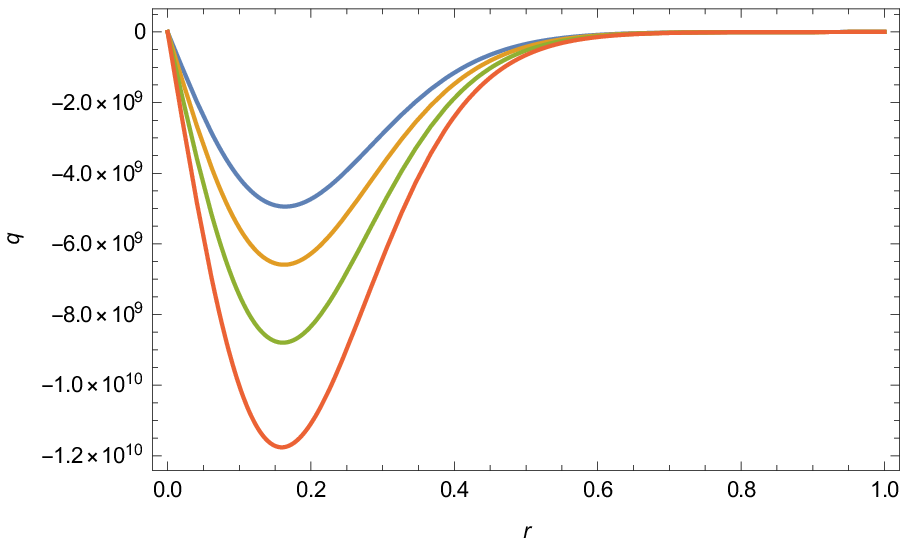}
\caption{\bf Evolution of heat flux with radial distance $r=0$ to $r=1$ for different values of time for $f(R) = R + \alpha R^2$ (top left), $f(R) = R^2$ (top right) and for $f(R) = e^{\alpha R}$ (bottom). Different curves in a single graph denotes evolution for different values of time; the uppermost curve denotes evolution for $t = 10.0$, the graph second from the top denotes evolution for $t = 10.2$ and so on. The lowermost graph denotes evolution for $t = 11.0$.}
\end{center}
\end{figure}

\begin{equation}
-2\frac{\dot{D}'}{D}+4\frac{\dot{D}D'}{D^2}= -\frac{qD^3}{F}+\frac{\dot{F}'}{F}-\frac{\dot{D}F'}{DF} -\frac{D'\dot{F}}{DF}.
\end{equation}

In Figure $6$, the evolution of heat flux is plotted as a function of $r$ for different values of time. The heat flux maintains a negative profile throughout the collapse. It vanishes both at the centre and at the boundary hypersurface. The radial distance from the centre at which the heat flux is maximum depends on the choice of $f(R)$. Different curves in a single graph denotes evolution for different values of time. The uppermost curve shows the distribution when $t = 10.0$ and the lowermost one is for $t = 11.0$. 

\section{Formation of a horizon}
The condition of the formation of an apparent horizon is given by the equation
\begin{equation}
g^{\mu \nu}{\it Y}_{,\mu} {\it Y}_{,\nu} = 0,
\end{equation}
where $Y(r, t)$ is the radius of the two-sphere, given by $r D(r, t)$ in the present case. From the evolution of the scale factor given by equation (\ref{exact}) one can rewrite this condition as
\begin{equation}
r^{2}\dot{\eta}^{2} - (\xi + \eta)^{2} - r^{2}\xi'^{2} + 2r\xi'(\xi + \eta) = 0.
\end{equation}
The time of formation of an apparent horizon is then given by
\begin{equation}
t_{ap} = t_{0} \pm \Big[\frac{8r_{b}}{3\xi_{1}}\Big(\xi - 2r \xi' + \frac{2\xi_{1}r_{b}}{9}- \xi_{0}\Big)\Big]^{\frac{1}{2}}.
\end{equation}
Since a collapsing scenario ($\dot{D} < 0$) is only ensured when $\xi_{1}$ is negative, and $r_{b}$ is always positive, a formation of horizon is possible if and only if $(\xi - 2r \xi' + \frac{2\xi_{1}r_{b}}{9}- \xi_{0}) > 0$. $\xi$ is given by 
\begin{equation}
 \xi(r) = \frac{4r_{b} r^6}{171\xi_1} - \frac{4r_{b}^3 r^4}{39\xi_1} + \frac{3}{14} r^{2} \Bigg(\frac{8r_{b}^5}{9\xi_1} + \frac{9\xi_1}{4r_{b}}\Bigg) - 3k_{1}r^{-\frac{1}{3}} +k_{2} .
\end{equation}

If this condition is satisfied, we actually have an example of a horizon without any singularity to cover, since the zero proper volume singularity is reached only asymptotically in comoving time, whereas the horizon forms in a finite comoving time.

\section{Evolution of the energy conditions}
In order that a collapsing fluid is physical, it must obey some energy conditions which we briefly discuss below.
\begin{enumerate}
\item {{\bf The Null Energy Condition} holds if for all null vectors
\begin{equation}
T_{\mu\nu} k^{\mu} k^{\nu} \geq 0
\end{equation}
}
\item {{\bf The weak energy condition} holds if for any non-spacelike vector $w^{\alpha}$,
\begin{equation}
T_{\alpha\beta} w^{\alpha} w^{\beta} \geq 0
\end{equation}
This is basically equivalent to saying that the energy density of the fluid is non-negative.
}
\item {{\bf The dominant energy condition} holds good if for a timelike vector $w^{\alpha}$, $- T_{\alpha\beta} w^{\beta}$ is timelike or null. This can be interpreted as a condition such that the speed of energy flow of matter is less than the speed of light for any observer.
}
\item {{\bf The strong energy condition} is satisfied if for any timelike unit vector $w^{\alpha}$
\begin{equation}
2 T_{\alpha\beta} w^{\alpha} w^{\beta} + T \geq 0,
\end{equation} 
where $T$ is the trace of the energy momentum tensor. The strong energy condition can only be violated if the energy density is negative or if there exists a large negative pressure component of the energy momentum tensor.
}
\end{enumerate}

The energy conditions were investigated for an imperfect fluid in rigorous details by Kolassis, Santos and Tsoubelis \cite{kola}, and Pimentel, Lora-Clavijo and Gonzalez \cite{pimentel} quite recently. 

\begin{figure}[h]
\begin{center}
\includegraphics[width=0.35\textwidth]{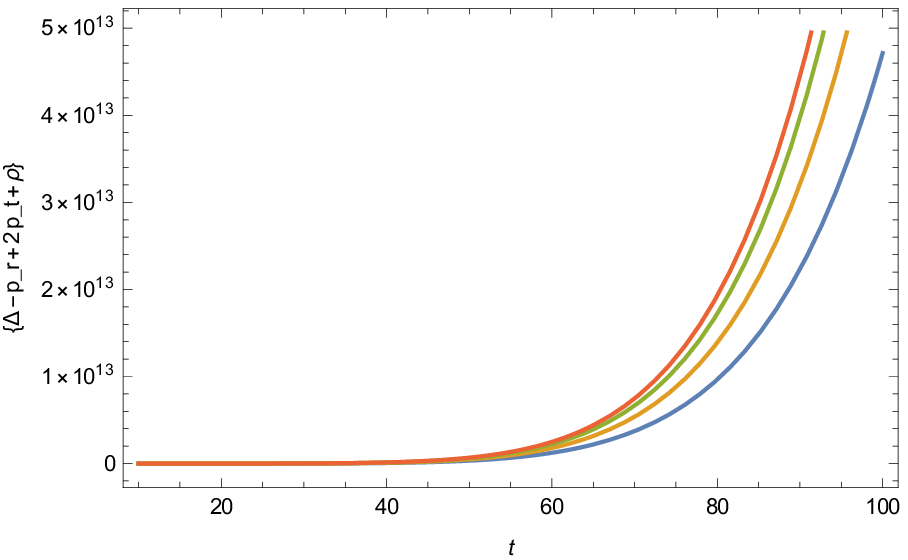}
\includegraphics[width=0.35\textwidth]{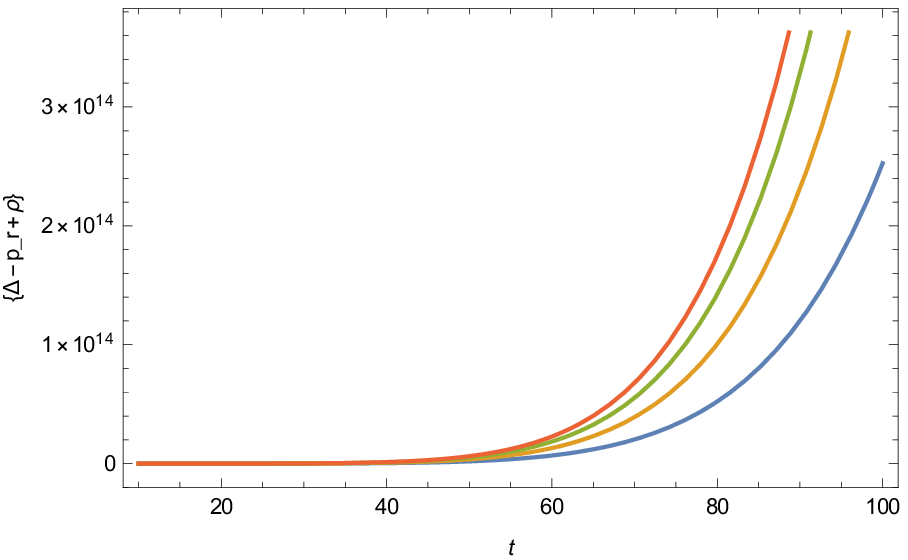}
\includegraphics[width=0.35\textwidth]{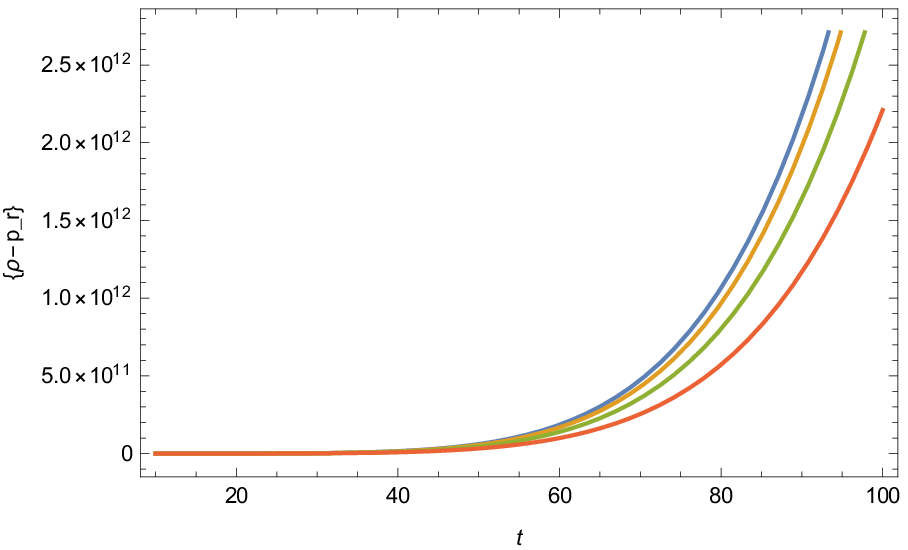}
\includegraphics[width=0.35\textwidth]{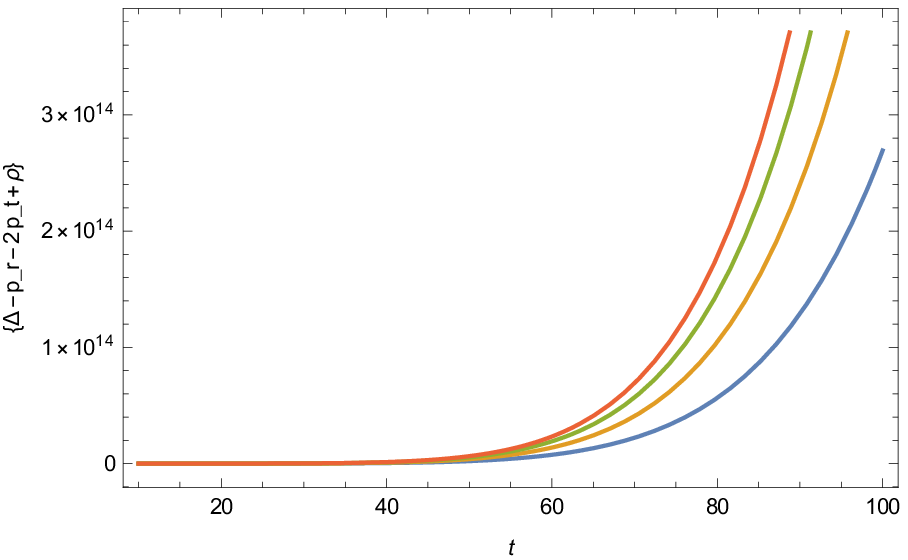}
\includegraphics[width=0.35\textwidth]{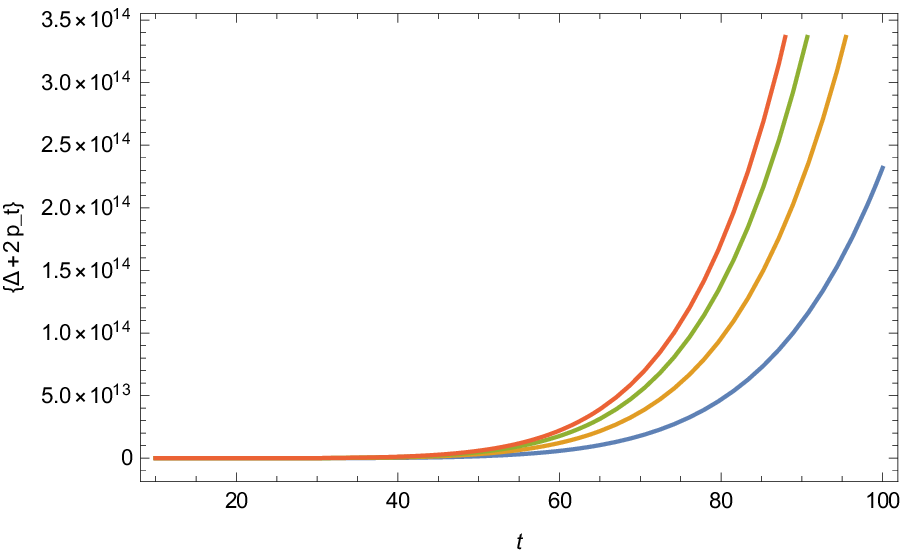}
\includegraphics[width=0.35\textwidth]{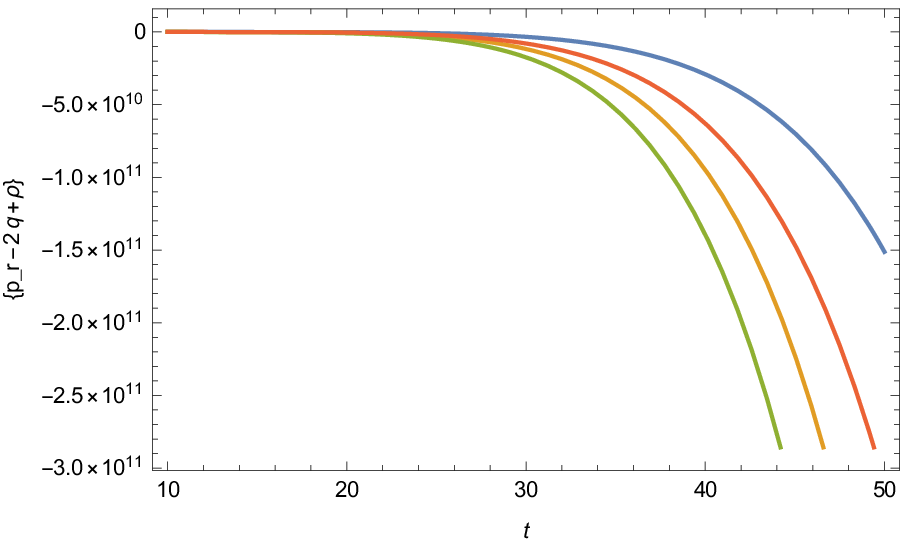}
\caption{\bf \small Evolution of the energy conditions with time, with different curves in a single graph denoting evolution for different radial distance from the centre for $f(R) = R + \alpha R^{2}$. Evolution of $NEC2$ (top left graph), $WEC$ (top right graph), $DEC1$ (middle left graph), $DEC2$ (middle right graph) and $SEC$ (bottom left graph) suggests that these energy conditions are valid throughout the collapse. For each of these graphs, the lowermost curve denotes the evolution when $r \sim 0$ and the topmost curve denotes the evolution when $r \sim r_{b}$. The $NEC1$ is given in the bottom right graph as a function of radial distance which is clearly violated throughout the course of the collapse.}
\end{center}
\end{figure}

Algebraically, the energy conditions to be satisfied are as follows,

{
\begin{enumerate}
\item{ (null energy conditions; $NEC1$ and $NEC2$) 
\begin{eqnarray}\label{ec1}
&\mid \rho + p_{r} \mid- 2\, \mid q\mid \geq 0 ~,\\
&\rho - p_{r} + 2\,p_{t}+\bigtriangleup \geq 0 ~,
\end{eqnarray}
}
\item{\textit{weak} energy conditions (WEC)  \\
\begin{equation}\label{ec2}
\rho - p_{r} +\bigtriangleup \geq 0 ~,
\end{equation}
}
\item{\textit{dominant} energy conditions ($DEC1$ and $DEC2$)  \\
\begin{eqnarray}\label{ec3}
&\rho - p_{r} \geq 0 ~, \\
&\rho - p_{r} -2\,p_{t} +\bigtriangleup \geq 0 ~,
\end{eqnarray}
}
\item{\textit{strong} energy conditions (SEC)  \\
\begin{equation}\label{ec4}
2\,p_{t}+ \bigtriangleup \geq 0~,
\end{equation}
}
\end{enumerate}
}
where $\bigtriangleup = \sqrt{(\rho + p_{r})^2 - 4\,q^2}$.

We plot the evolution of the energy condition with respect to time $t$ for different values of radial coordinate $r$, ranging from the centre at $r = 0$ to the boundary at $r = 1$. 

\begin{figure}[h]
\begin{center}
\includegraphics[width=0.35\textwidth]{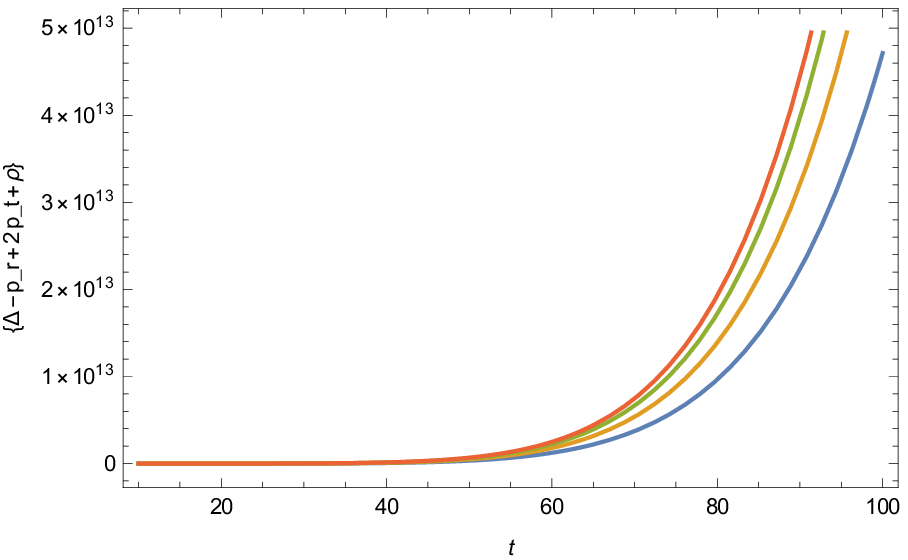}
\includegraphics[width=0.35\textwidth]{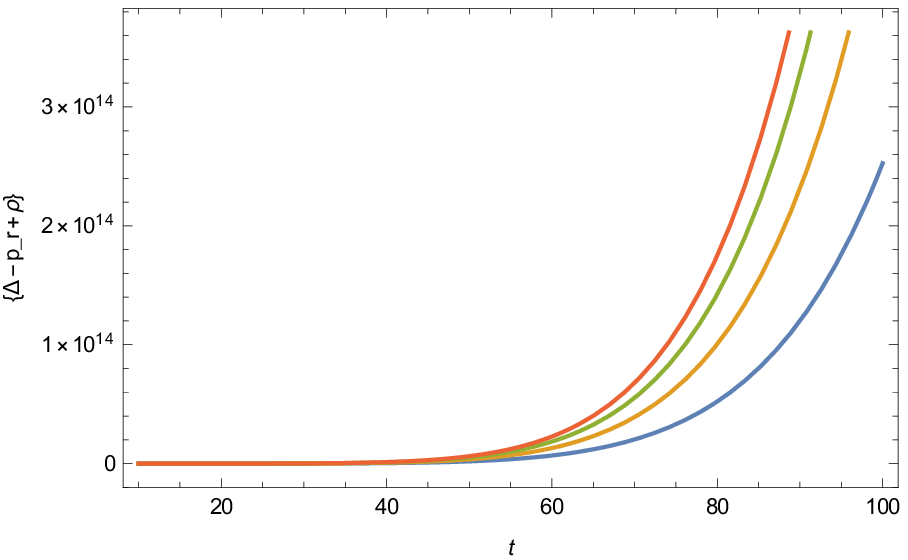}
\includegraphics[width=0.35\textwidth]{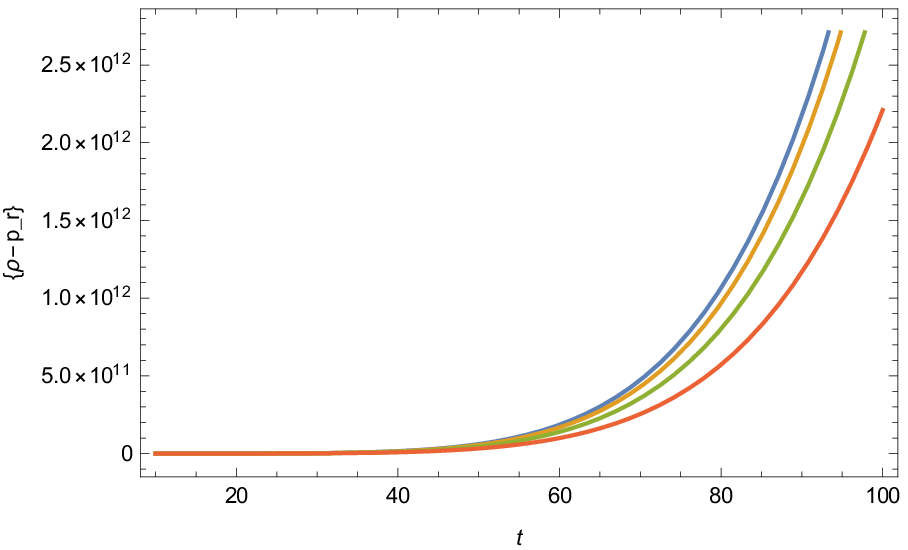}
\includegraphics[width=0.35\textwidth]{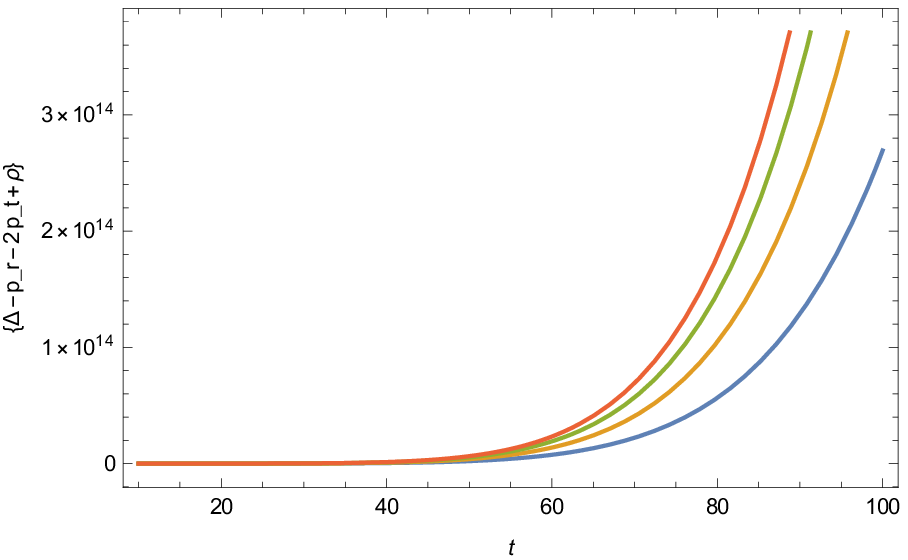}
\includegraphics[width=0.35\textwidth]{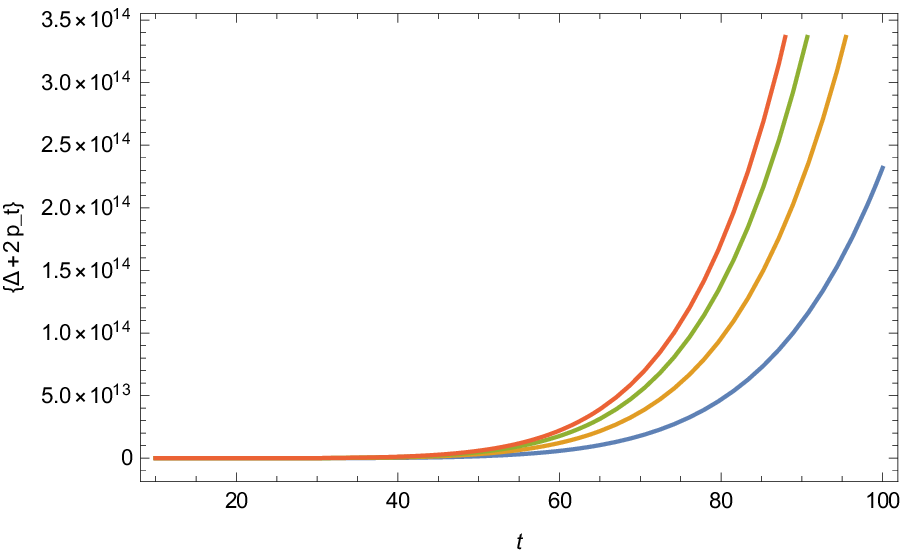}
\includegraphics[width=0.35\textwidth]{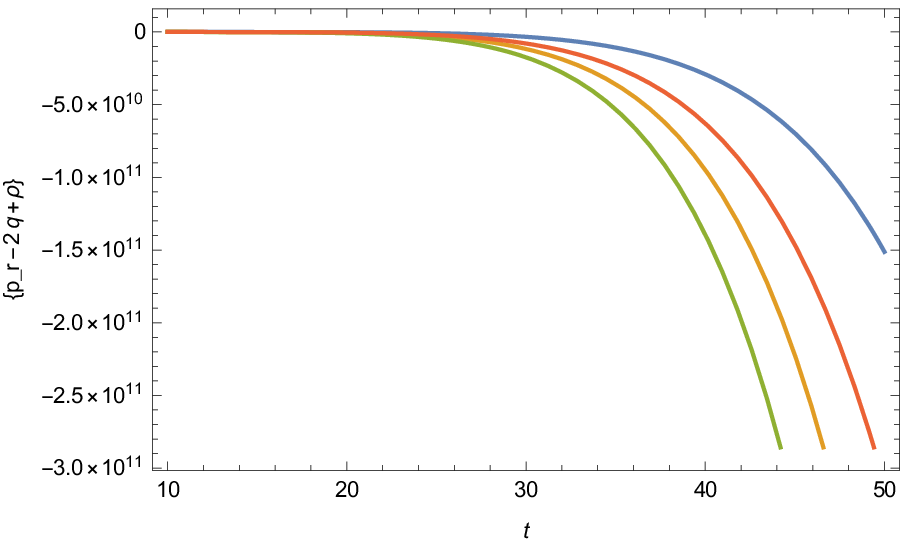}
\caption{\bf \small Evolution of the energy conditions with time, with different curves in a single graph denoting evolution for different radial distance from the centre for $f(R) = R^{2}$. Evolution of $NEC2$ (top left graph), $WEC$ (top right graph), $DEC1$ (middle left graph), $DEC2$ (middle right graph) and $SEC$ (bottom left graph) suggests that these energy conditions are valid throughout the collapse. For each of these graphs, the lowermost curve denotes the evolution when $r \sim 0$ and the topmost curve denotes the evolution when $r \sim r_{b}$. The $NEC1$ is given in the bottom right graph as a function of radial distance which is clearly violated throughout the course of the collapse.}
\end{center}
\end{figure}

We now investigate the evolution of energy conditions for a few choice of $f(R)$. We include a model with $f (R) = R + \alpha R^2$ in our study. As discussed by Clifton, Ferreira, Padilla and Skordis, such an $f(R)$ model describes well the temperature anisotropies in Cosmic Microwave Background \cite{padilla}. Presence of a term $\alpha R^2$ serves even more purpose, for instance, matter instability and difficulties regarding local gravity constraints in models like $f (R) \sim R - \frac{\alpha}{R^n}$ can be solved by inclusion of an $R^2$ term, as discussed by Paul, Debnath and Ghoshe \cite{paul}. Addition of an $R^2$ term into the action can also lead to consistent $f(R)$ model that passes the solar system tests, bypassing instability problems as shown by Nojiri and Odintsov \cite{nojiodi1, nojiodi2}. Since we are studying a scenario of gravitational collapse rather than a cosmological one, we also study the evolutions for a couple of $f(R)$ models which grow rapidly with $R$ at high curvature ($f(R) = R^2$ and $f(R) = e^{\alpha R}$). We also study a case where $f(R) \sim R^{1 + \delta}$, where $\delta$ is a preassigned small quantity. Clifton and Barrow studied the weak-field properties of such a model and found that it is compatible with local astronomical tests (e.g. perihelion shift), iff the $f (R)$ lagrangian is very close to General Relativity (in $f (R) \sim R^{(1+\delta)}$, $\delta$ must be of the order of $10^{-19}$ or less \cite{cliffbarrow}). \\

We make a note here that a choice of $f(R) \sim R^2$ is arguable since this theory does not have a proper Newtonian limit as discussed by Pechlaner and Sexl \cite{pech} in their discussion on quadratic lagrangians in general relativity. Moreover, an $f(R)$ model is compatible with local astronomical tests only if the lagrangian is very close to general relativity (for $f (R) \sim R^{(1 + \delta)}$ , $\delta$ can be atmost of the order of $10^{-19}$ \cite{cliffbarrow}). However, we keep the case of $f(R) \sim R^2 $ atleast for the sake of illustration.

{\bf \large $1. f(R) = R + \alpha R^2$} \\

Evolution of the energy conditions with time is shown in figure $7$ for $f(R) = R + \alpha R^{2}$ with $\alpha \sim 10^{-3}$. Different curves in a single graph denotes the evolution for different radial distance from the centre. Evolution of $NEC2$ (top left graph), $WEC$ (top right graph), $DEC1$ (middle left graph), $DEC2$ (middle right graph) and $SEC$ (bottom left graph) shows that they are positive throughout the collapse and therefore, holds true. For each of these graphs, the lowermost curve denotes the evolution near the centre of the collapsing star ($r \sim 0$) and the topmost curve denotes the evolution close to the boundary ($r \sim r_{b}$). The $NEC1$ is given in the bottom right graph as a function of radial distance which is negative and is clearly violated throughout the course of the collapse. \\

\begin{figure}[h]
\begin{center}
\includegraphics[width=0.35\textwidth]{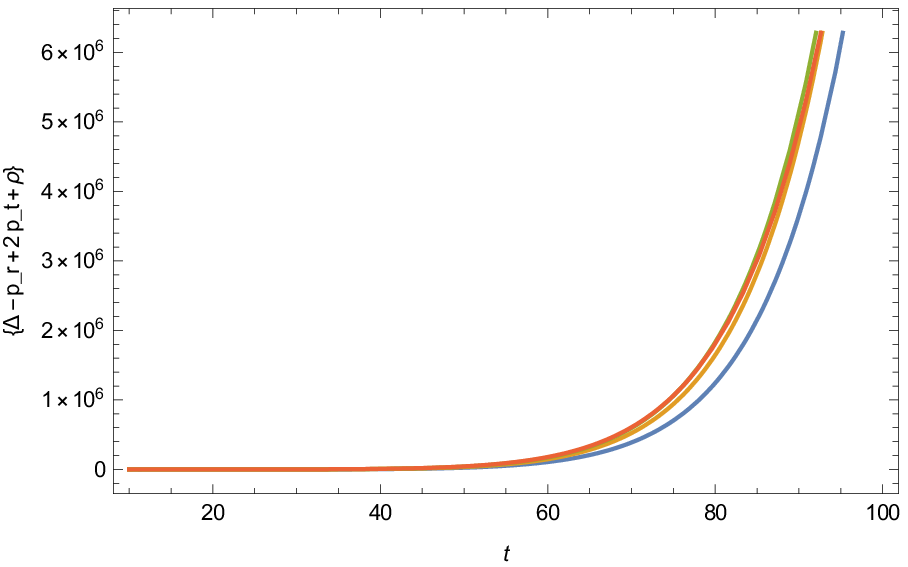}
\includegraphics[width=0.35\textwidth]{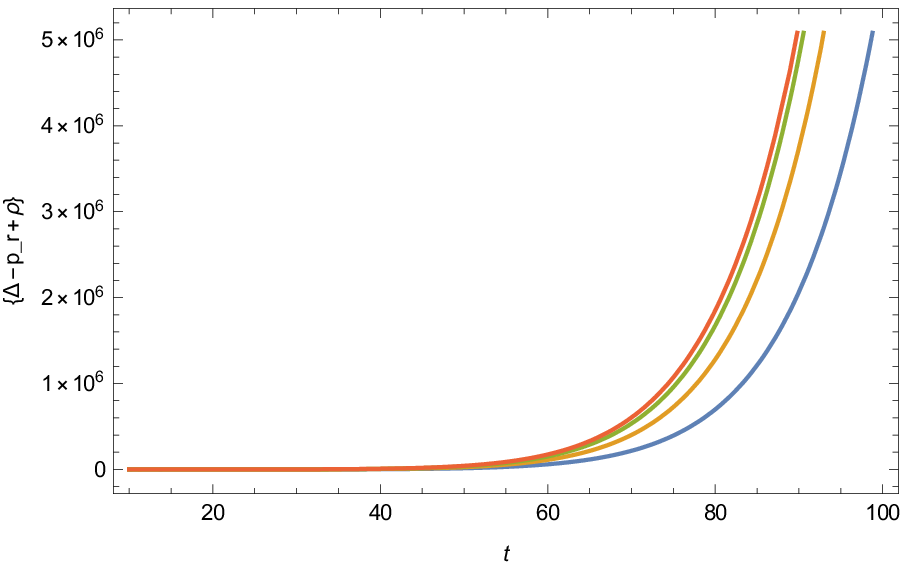}
\includegraphics[width=0.35\textwidth]{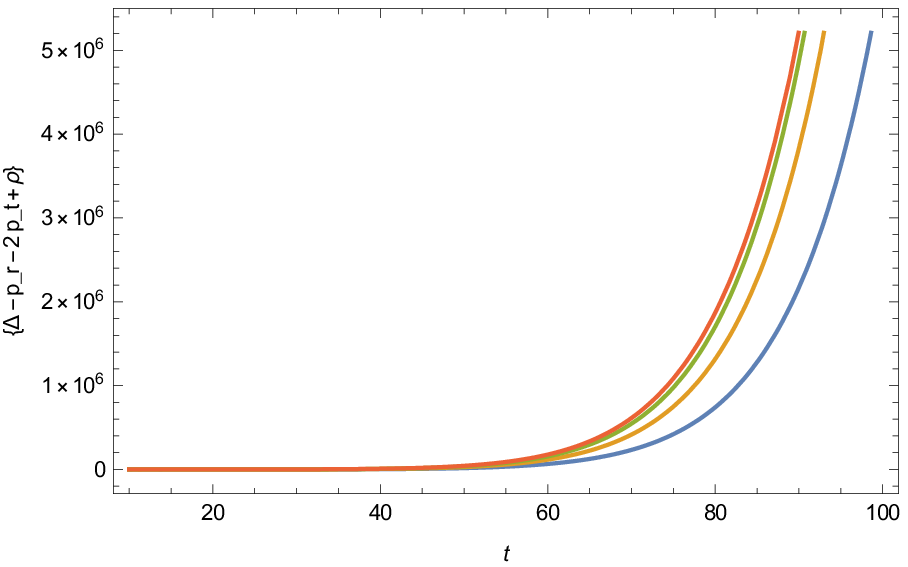}
\includegraphics[width=0.35\textwidth]{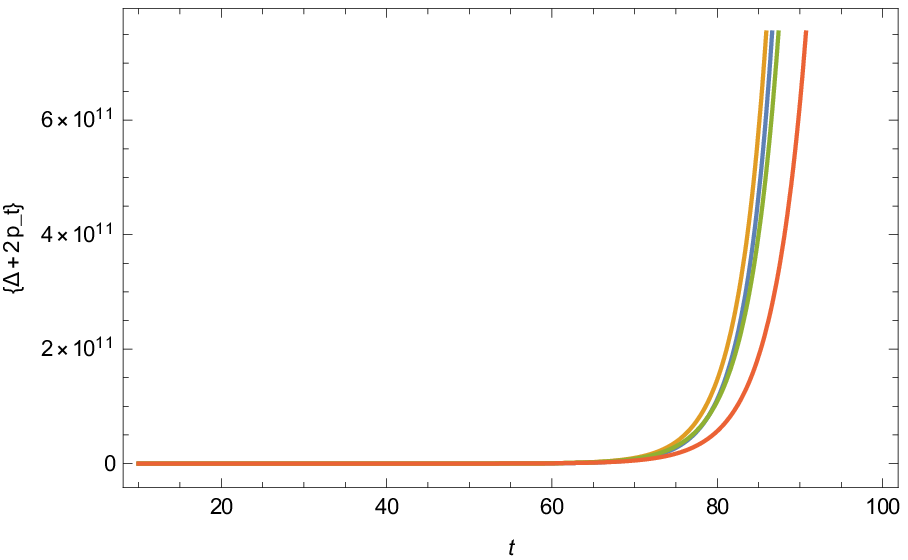}
\caption{\bf \small Evolution of the energy conditions with time, with different curves in a single graph denoting evolution for different radial distance from the centre for $f(R) = e^{\alpha R}$. Evolution of $NEC2$ (top left graph), $WEC$ (top right graph), $DEC2$ (bottom left graph) and $SEC$ (bottom right graph) suggests that these energy conditions are valid throughout the collapse. For each of these graphs, the lowermost curve denotes the evolution when $r \sim 0$ and the topmost curve denotes the evolution when $r \sim r_{b}$.}
\end{center}
\end{figure}

{\bf \large $2. f(R) = R^2$} \\

Evolution of the energy conditions with time is shown in figure $8$ for $f(R) = R^{2}$. Different curves in a single graph denotes the evolution for different radial distance from the centre. Evolution of $NEC2$ (top left graph), $WEC$ (top right graph), $DEC1$ (middle left graph), $DEC2$ (middle right graph) and $SEC$ (bottom left graph) shows that they are positive throughout the collapse and therefore, holds true. For each of these graphs, the lowermost curve denotes the evolution near the centre of the collapsing star ($r \sim 0$) and the topmost curve denotes the evolution close to the boundary ($r \sim r_{b}$). The $NEC1$ is given in the bottom right graph as a function of radial distance which is negative and is clearly violated throughout the course of the collapse. \\

{\bf \large $3. f(R) = e^{\alpha R}$} \\

Evolution of the energy conditions with time for $f(R) = e^{\alpha R}$ are plotted in Figure $9$ and $10$. Different curves in a single graph denotes the time evolution for different radial distance from the centre. Evolution of $NEC2$ (top left graph of Figure $9$), $WEC$ (top right graph of Figure $9$), $DEC2$ (bottom left graph of Figure $9$) and $SEC$ (bottom right graph of Figure $9$) suggests that these energy conditions are strictly positive and valid throughout the collapse. The lowermost curve for each of these graphs denotes the evolution when $r \sim 0$ and the topmost curve denotes the evolution when $r \sim r_{b}$. \\
For $f(R) = e^{\alpha R}$, the evolution of $NEC1$ is shown on the left graph of Figure $10$. The lowermost curve denotes the evolution when $r \sim 0$, i.e., close to the centre, where the energy condition is clearly violated. However, the evolution very close to the boundary is denoted by the topmost curve which maintains a positive profile throughout the collapse, i.e., the energy condition is valid close to the boundary hypersurface $r \sim r_{b}$. A similar evolution is seen for the evolution of $DEC1$ which is plotted on the right graph of Figure $10$. One can infer from this result that even if all the energy conditions are well-satisfied at the boundary surface of the dying star, it does not always guarrantee the validity of all the energy conditions in as one approaches the centre. \\

\begin{figure}[h]
\begin{center}
\includegraphics[width=0.40\textwidth]{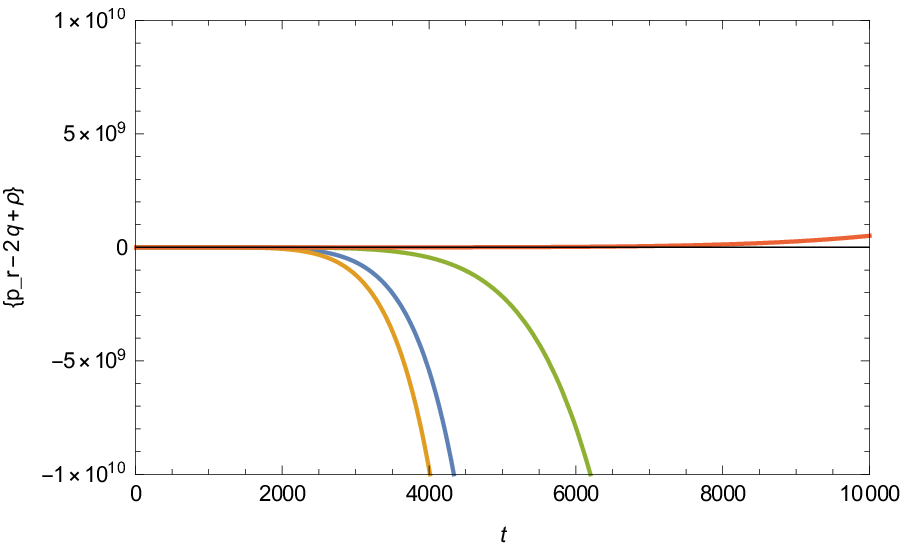}
\includegraphics[width=0.40\textwidth]{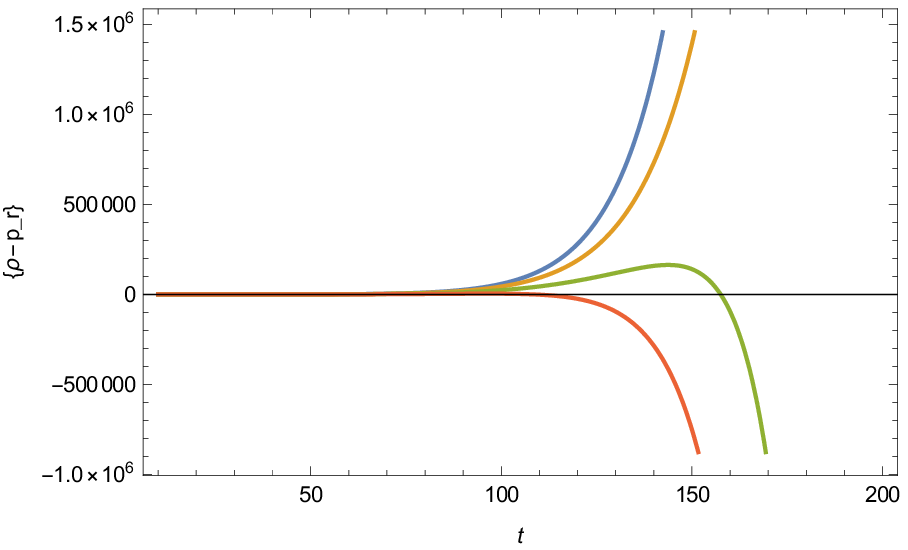}
\caption{\bf \small Evolution of the energy conditions with time, with different curves in a single graph denoting evolution for different radial distance from the centre for $f(R) = e^{\alpha R}$. Evolution of $NEC1$ is shown on the left graph. The lowermost curve denotes the evolution when $r \sim 0$, i.e., close to the centre, where the energy condition is clearly violated. However, the evolution very close to the boundary is denoted by the topmost curve which maintains a positiove profile throughout the collapse, i.e., the energy condition is valid close to the boundary hypersurface $r \sim r_{b}$. A similar evolution is seen for the evolution of $DEC1$ which is plotted on the right graph.}
\end{center}
\end{figure}

{\bf \large $4. f(R) = R^{(1+ \delta)}$} \\

Evolution of the energy conditions with time is shown in figure $11$ for an $f(R)$ model for which the departure of the $f(R)$ modification from Einstein's gravity is very small. This is given by $f(R) = R^{(1+ \delta)}$, where $\delta$ is a very small positive number, $\sim 10^{-19}$ or lower. Different curves in a single graph denotes the evolution for different radial distance from the centre. Evolution of $NEC2$ (top left graph), $WEC$ (top right graph), $DEC1$ (middle left graph), $DEC2$ (middle right graph) and $SEC$ (bottom left graph) shows that they are positive throughout the collapse and therefore, holds true. For each of these graphs, the lowermost curve denotes the evolution near the centre of the collapsing star ($r \sim 0$) and the topmost curve denotes the evolution close to the boundary ($r \sim r_{b}$). The $NEC1$ is given in the bottom right graph as a function of radial distance which is negative and is clearly violated throughout the course of the collapse. These behavior is in fact, qualitatively similar as that of an $R^2$ model. \\

\begin{figure}[h]
\begin{center}
\includegraphics[width=0.35\textwidth]{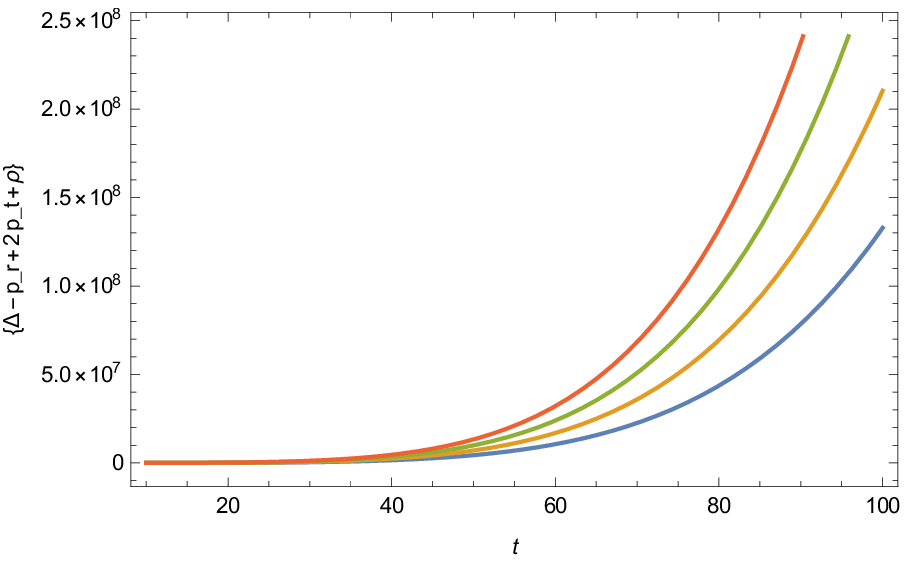}
\includegraphics[width=0.35\textwidth]{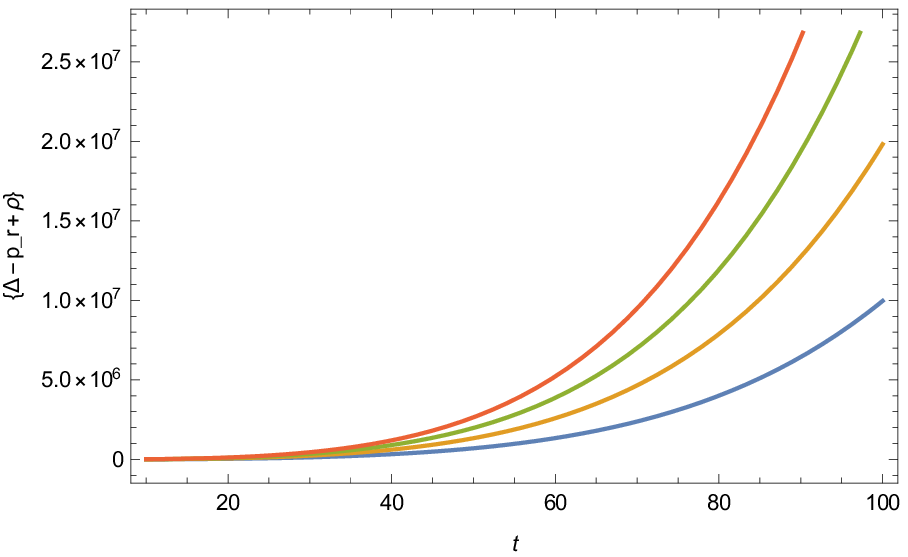}
\includegraphics[width=0.35\textwidth]{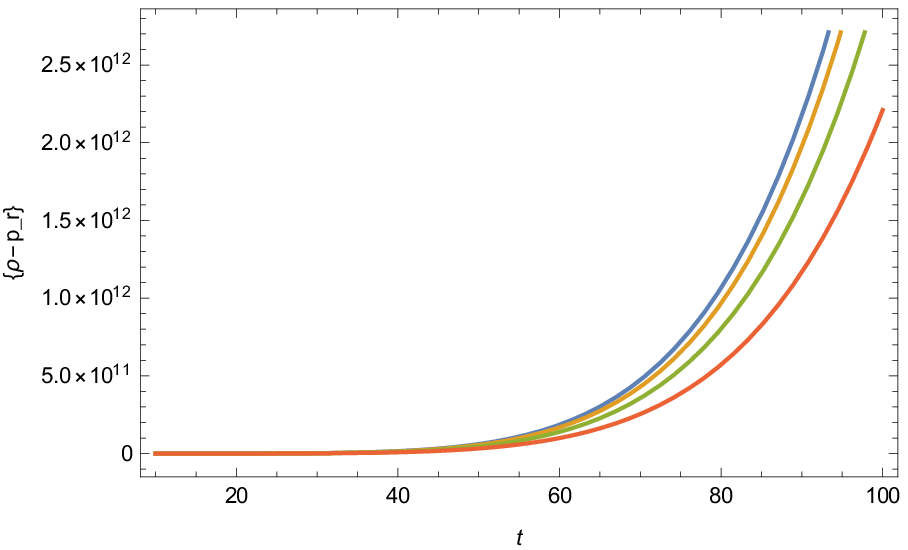}
\includegraphics[width=0.35\textwidth]{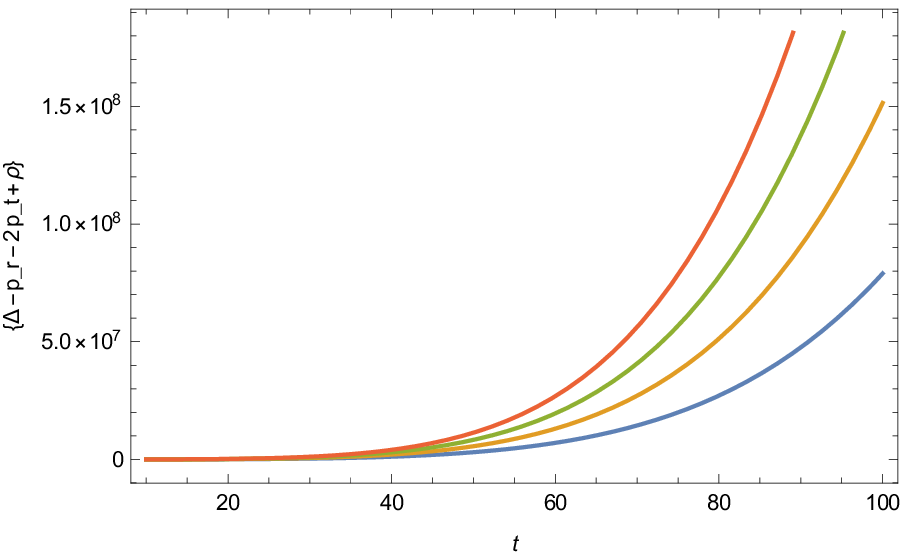}
\includegraphics[width=0.35\textwidth]{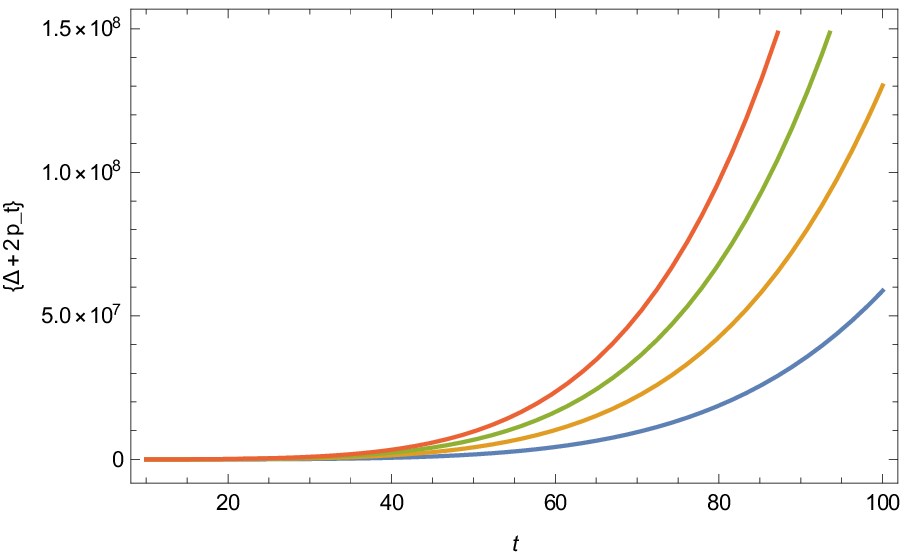}
\includegraphics[width=0.35\textwidth]{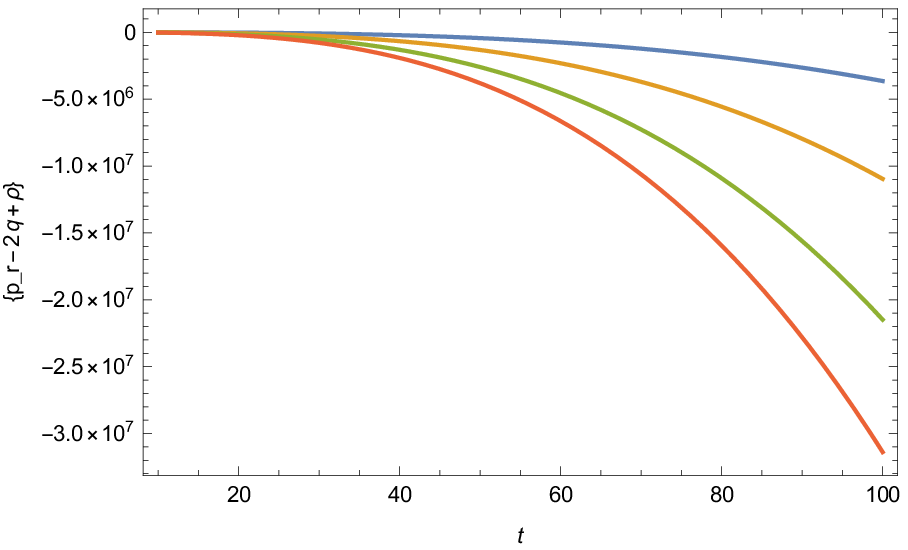}
\caption{\bf \small Evolution of the energy conditions with time, with different curves in a single graph denoting evolution for different radial distance from the centre for $f(R) = R^{(1+ \delta)}$. Evolution of $NEC2$ (top left graph), $WEC$ (top right graph), $DEC1$ (middle left graph), $DEC2$ (middle right graph) and $SEC$ (bottom left graph) suggests that these energy conditions are valid throughout the collapse. For each of these graphs, the lowermost curve denotes the evolution when $r \sim 0$ and the topmost curve denotes the evolution when $r \sim r_{b}$. The $NEC1$ is given in the bottom right graph as a function of radial distance which is clearly violated throughout the course of the collapse.}
\end{center}
\end{figure}

The energy conditions are actually some constraints to be followed such that the notion of locally positive energy density holds everywhere. In the present case, some are indeed violated. But one must realize that this is not really a serious infringement on the physical plausibility of the system. First of all, the positivity of energy is never violated, so the system is quite well posed for a quantization scheme. Secondly, the energy conditions that we derived are not for the fluid alone, it takes care of the contribution of curvature to the effective stress-energy tensor as well. The violation of some of the energy conditions can be attributed to the contribution from curvature, which does not really have the onus of satisfying the energy conditions. There are lots of examples of quite reasonable classical systems, which can violate some of the energy conditions. The simplest example is a massive scalar field non-minimally coupled to gravity, as discussed by Visser and Barcelo \cite{visser, barcelo}. One interesting outcome of the solutions of gravitational field equations that disobey energy conditions is that they can give rise to Lorentzian wormholes as discussed by Barcelo \cite{barcelo}, Flanagan and Wald \cite{flanagan}.   

\section{Conclusion}
We investigated a collapsing spherical star in a very general setup of $f(R)$ modifications of gravity, for a conformally flat spacetime. It was discussed quite recently that the smooth matching of a collapsing interior with a vacuum exterior is not at all a straightforward task in the context of $f(R)$ theories \cite{goswamicollapse}. The extra matching conditions on the Ricci scalar enforce strong restrictions on the allowed solutions. The present work utilizes these conditions required to close the system of equations. It deserves mention that some classes of collapsing models allowed in general relativity are in fact ruled out in $f(R)$ gravity models \cite{goswamicollapse}. The solutions presented here satisfy all the boundary matching conditions and they describe the interior of a collapsing star with pressure anisotropy and heat flux, without any assumption of equation of state or any particular choice of $f(R)$ at the outset. It is quite interesting to note that the interior distribution needs to be inhomogeneous for the proper matching at the boundary. However, whether the matter distribution of the star satisfies all the reasonable energy conditions or not, depends on the choice of functional form of $f(R)$.  \\

We note here that the system under consideration in the manuscript is chosen to be conformally flat at the outset. This is indeed a very special situation, in the sense that there remains only one scalar gravitational degree of freedom, namely the conformal factor $D$. This might have been responsible for a loss of generality in the collapsing evolution. However, the high non-linearity of the fourth order field equations infact compels one to resort to a simplified system which can describe the overall physics nonetheless. The results in the manuscript therefore may not represent a complete general spherical gravitational collapse, but are physically relevant and interesting as an example itself. We also note that a Schwarzschild solution can be an exact solution for $f (R)$ theories only when the curvature becomes constant. From the constant curvatre condition $R f'(R) = 2 f(R)$, one can argue that the exponential $f (R)$ cases may not produce a constant curvature solution at all. This is not too unexpected either, since in the presence of a higher curvature term in the lagrangian of gravity the smooth matching of the interior with an exterior Schwarzschild metric may be spoilt sometimes, as discussed by Banerjee and Paul \cite{nbtp}, Chakrabarti very recently \cite{sc}.

The collapsing star reaches a zero proper volume only asymptotically, since the rate of collapse dies down with time. Therefore, whether the collapse ends up in a singularity or not becomes actually irrelevant. Recent studies of conformally flat collapsing solutions have shed light on many different possible outcomes of a gravitational collapse. For instance, a conformally flat collapsing perfect fluid was studied by Hamid, Goswami and Maharaj \cite{hamid} quite recently. It was proved that if the strong energy condition is obeyed, the collapse necessarily leads to a black hole end-state. A spatially homogeneous massive scalar field collapse also leads to a black hole end-state under conformal flatness, as discussed by Chakrabarti and Banerjee \cite{scnbscalar1}. It was also proved that in the presence of pressure anisotropy and a dissipation, the spacetime singularity can remain naked \cite{scnbscalar2}. In the present case, however, it is obvious that for a similar collapsing body in $f(R)$ gravity, the end state may not end up in either a black hole or a naked singularity afterall. \\

There are quite a few examples of specific $f(R)$ gravity models where a spherical collapse to a singularity at a finite future is possible \cite{scnb1, scnbeuro}. In the present case interior inhomogeneity leads to an acceleration unlike the standard Tolman-Bondi collapsing solutions in general relativity. The divergence of this acceleration and the dissipative effects like the heat conduction should be the key to this infinite delay in the formation of the singularity as expected from the Raychaudhuri equation \cite{akr}.

\section{Acknowledgements}
N.B. wishes to thank the Astrophysics and Cosmology Research Unit at University of Kwazulu-Natal for a warm hospitality, where the work was initiated. S.C. wishes to thank Prof. Sayan Kar for useful discussions. S.C. was supported by the National Post-Doctoral Fellowship (file number: PDF/2017/000750) from the Science and Engineering Research Board (SERB), Government of India.

\end{document}